\documentclass[pra,amsfonts,floatfix,superscriptaddress,twocolumn]{revtex4-2}
\usepackage[utf8]{inputenc}
\usepackage{amssymb,amsmath}
\usepackage{hyperref, float}
\usepackage{graphicx}% Include figure files
\usepackage{tikz}
\usepackage{tikz-cd}
\usetikzlibrary{%
  matrix,%
  calc,%
  arrows%
}

\newcommand{\cM}{{\mathcal{M}}}

\newcommand{\cQ}{{\mathcal{Q}}}
\newcommand{\cX}{{\mathcal{X}}}
\newcommand{\cY}{{\mathcal{Y}}}
\newcommand{\cH}{{\mathcal{H}}}
\newcommand{\bR}{{\mathbb{R}}}
\newcommand{\bZ}{{\mathbb{Z}}}

\DeclareMathOperator{\Diff}{Diff}

\begin{document}

\title{Topological Exchange Statistics in One Dimension}

\author{N.L.  Harshman}
\affiliation{American University, Washington DC, USA}
\author{A.C.  Knapp}
\affiliation{University of Florida, Gainesville, FL, USA}

\begin{abstract}
  The standard topological approach to indistinguishable particles formulates exchange statistics by using the fundamental group to analyze the connectedness of the configuration space.  Although successful in two and more dimensions, this approach gives only trivial or near trivial exchange statistics in one dimension because two-body coincidences are excluded from configuration space.  Instead, we include these path-ambiguous singular points and consider configuration space as an orbifold.  This orbifold topological approach allows unified analysis of exchange statistics in any dimension and predicts novel possibilities for anyons in one-dimensional systems, including non-abelian anyons obeying alternate strand groups.  These results clarify the non-topological origin of fractional statistics in one-dimensional anyon models.
\end{abstract}

\maketitle

\section{Introduction}

Although all fundamental particles and most composite particles either satisfy Bose-Einstein (BE) or Fermi-Dirac (FD) statistics, recent measurements provide the strongest evidence yet that the quasiparticle excitations in the fractional quantum Hall effect (FQHE) obey \emph{fractional exchange statistics} \cite{bartolomei_fractional_2020, nakamura_direct_2020}.  Wilczek coined the term anyons for particles that obey fractional exchange statistics~\cite{wilczek_quantum_1982} and these recent experiments demonstrate that the excitations in the FQHE characterized by a $\nu=1/3$ filling factor exhibit a $\theta = \pi/3$ exchange phase, satisfying a prediction made nearly forty years ago~\cite{halperin_statistics_1984, arovas_fractional_1984, wilczek_fractional_1990}.

The origin of fractional exchange phases in the FQHE can be traced to a topological peculiarity: for two-dimensional systems with two-body coincidences excluded, the configuration space is not simply-connected~\cite{leinaas_theory_1977}.  Not all paths in configuration space that exchange particles are equivalent, and the equivalence classes of possible exchange paths are described by the braid group.  Abelian representations of the braid group are characterized by an arbitrary exchange phase $\theta \in [0,2\pi)$ that determines the fractional exchange statistics parameter $\theta$~\cite{wu_general_1984, wu_multiparticle_1984, forte_quantum_1992, khare_fractional_2005}.  Because of their topological origin, the fractional exchange statistics of these anyons can be `transmuted' into a gauge interaction, i.e.~the charged flux-tube model~\cite{wilczek_magnetic_1982, khare_fractional_2005}. Quasiparticle excitations obeying non-abelian braid group statistics may also exist in the FQHE~\cite{moore_nonabelions_1991, wen_non-abelian_1991, nayak_non-abelian_2008}, but unambiguous confirmation remains experimentally elusive.  Adiabatic exchanges of non-abelian anyons could provide an implementation for fault-tolerant quantum computing \cite{kitaev_fault-tolerant_2003, freedman_topological_2003}.  This application, combined with a fundamental interest in understanding topological states of matter, continues to drive interest in engineering novel physical systems that support excitations with non-standard exchange statistics \cite{nayak_non-abelian_2008, alicea_non-abelian_2011, maciazek_non-abelian_2019}. 

Can the topological approach to exchange statistics in two-dimensional systems be applied to one-dimensional systems? There is debate in the literature.  Unlike in two or more dimensions, particles in one-dimensional systems must pass through each other to exchange positions and even short range (or zero-range) interactions have dramatic dynamic and thermodynamic effects~\cite{minguzzi2022strongly}.  Given the inevitable intermingling of interactions with exchange in one dimension, can a purely topological exchange phase be separated from the dynamical phase accumulated along the exchange trajectory~\cite{ha_fractional_1995}? Further, for indistinguishable particles, what does ``pass through each other'' even mean? The two-body coincidence is a singular point in configuration space that introduces ambiguity~\cite{bourdeau_when_1992}, so how can one distinguish trajectories in which two particles reflect from those in which they transmit?

The standard formulation of topological exchange statistics is not sufficient as it excludes two-body coincidences from configuration space.  In one dimension, the removal of two-body coincidences makes particle exchanges impossible, and so the standard formulation of topological exchange statistics allows only trivial representations~\cite{polychronakos_generalized_1999, nayak_non-abelian_2008}.  To overcome this technical limitation, we extend the standard formulation of topological exchange statistics by treating the configuration space of indistinguishable particles as an \emph{orbifold}~\cite{bourdeau_when_1992, landsman_quantization_2016, ohya_generalization_2021}, a generalization of the idea of a manifold~\cite{thurston_geometry_2002, adem_orbifolds_2007}.  Informally, an orbifold is locally equivalent to the linear quotient of a Euclidean space by a finite group and `remembers' that symmetry.  This extension allows trajectories that reflect and transmit at two-body coincidences to be topologically distinguished by elements of the \emph{orbifold fundamental group}, even for indistinguishable particles.

We present details about orbifolds in Sect.~II, but a simple example of an orbifold is the configuration space for two indistinguishable particles on a line, equal to a plane modulo a reflection, $\bR^2/S_2$.  This quotient identifies indistinguishable configurations $(x_1,x_2)$ and $(x_2,x_1)$ in $\bR^2$ and has singular locus $x_1 = x_2$.  In $\bR^2$, singular points like $(x,x)$ have a Euclidean neighborhood, but how do they look in the quotient space? One could collapse all identified points to obtain the underlying space, the half-plane $|\bR^2/S_2|$, as a manifold with boundary~\cite{leinaas_theory_1977}.  Alternatively, one could consider $\bR^2/S_2$ as an orbifold, a two-dimensional space with an internal edge of orbifold singularities on the reflection line.  To elucidate the difference, consider the perspective of an ant sitting on the singular locus of $\bR^2/S_2$.  In the half-plane perspective, there is an ``edge of the universe'' which no path can cross.  The half-plane is simply-connected and there are no topological exchange statistics.  However, in the orbifold perspective, the ant sees itself as sitting in the middle of an infinite plane where the view just happens to be symmetric.  Paths that cross (transmit) and do not cross (reflect) this internal edge can be distinguished by the local observer.  The orbifold fundamental group in this case is the symmetric group $S_2$, giving the possibility for both bosonic and fermionic topological exchange statistics.

%In Sect.~II we provide an overview of the orbifold approach.  We show that, the orbifold extension to the topological approach resolves the ambiguity of singular points when indistinguishable particles collide and provides a unified approach to particle systems in any dimension encapsulated in Eq.~(\ref{eq:gensym}).  Sect.~II is supplemented by two appendices that give further background on the topological approach to quantization and to exchange statistics.

We first introduced the orbifold extension to the topological approach to exchange statistics in \cite{harshman_anyons_2020}.  This analysis was motivated by the observation that, like two-body interactions in two dimensions, three-body interactions in one dimension cause co-dimension two defects in configuration space.  Engineering such interactions could be feasible in ultracold atomic gases in optical traps and optical lattices \cite{buchler_three-body_2007, daley_effective_2014, mahmud_dynamically_2014, valiente_three-body_2019}.  The orbifold fundamental group  arising from hard-core three-body interactions in one dimension is similar to the braid group: it is an infinite discrete group that descends from the symmetric group by breaking one of the generating relations. Further, it is realized by strand diagrams obeying certain crossing rules.  We named it the \emph{traid} group by analogy, but mathematicians also had discovered it in other contexts and given it various other names~\cite{cisneros2020alexander}, including the doodle group~\cite{khovanov_doodle_1997, bartholomew_doodles_2016}, the planar braid group~\cite{gonzalez_linear_2021}, and the twin group~\cite{bardakov_structural_2019, Naik20}.

We extend these results in Sect.~III and classify all possible topological exchange statistics for distinguishable and indistinguishable particles in one dimension.  We consider the only two topological types of path-connected one-dimensional base manifolds, the interval-type (which includes the infinite interval of the real line) and circle, and we analyze all possible topologically non-trivial interactions.  This includes hard-core two-body interactions, three-body interactions, and one other non-trivial form: a non-local four-body interaction for which pairwise coincidences are allowed, but not pairs of pairwise coincidences.  This interaction leads to another descendent of the symmetric group that we call the \emph{fraid} group. It is realized by strand diagrams with non-local relations between the generators.

Our results in Sect.~IIIB demonstrate that on the twisted configuration space appropriate for indistinguishable particles on a ring, the possibility exists for non-abelian generalized parastatistics for soft-core particles.  We also extend the traid and fraid group (and their combination) to the ring geometry and discuss their `pure' forms that apply to distinguishable particles.

Although a topological definition for exchange statistics in one dimension did not previously exist, there is a large literature of one-dimensional models where the particles are `anyons' and/or have `fractional statistics'.  Continuum models include (references are representative, not exhaustive)~\footnote{Our analysis does not consider the anyon-Hubbard model~\cite{amico_one-dimensional_1998, keilmann_statistically_2011, hao_dynamical_2012, wright_nonequilibrium_2014, greschner_2015, straeter_2016, lange_strongly_2017}  because our topological methods do not extend to configuration spaces with discrete topology, although we note that the continuum limit of the anyon-Hubbard model has recently been derived~\cite{bonkoff_bosonic_2021}.}: (1) Leinaas-Myrheim anyons~\cite{leinaas_theory_1977, HANSSON1992559, posske_2017}; (2) Calogero-Sutherland anyon models~\cite{ha_fractional_1995, polychronakos_generalized_1999, SREERANJANI20091176}; (3) hard-core anyon models with $\delta$-type interactions and fractional exchange statistics~\cite{zhu_topological_1996, girardeau_anyon-fermion_2006, delcampo_fermionization_2008}; and (4) Kundu/Lieb-Liniger anyons with non hard-core $\delta$-type interactions and fractional exchange statistics~\cite{kundu_exact_1999, batchelor_one-dimensional_2006, Patu_2007, hao_dynamical_2012, zinner_strongly_2015, posske_2017, stouten_something_2018, patu_correlation_2019, valiente_bose-fermi-2020, bonkoff_bosonic_2021, valiente_universal_2021}.  A full survey of these model exceeds the scope of this article (for a brief review see Appendix A of Ref.~\cite{posske_2017}), but a unifying property of these models is that they contain a parameter that interpolates between bosonic and fermionic limits. In Sect.~IV, we give a preliminary analysis of these four classes of continuum models and conclude that any braid-like exchange phases in models (2)-(4) originate from dynamics, i.e., they are non-topological in origin and cannot be absorbed into a statistical gauge interaction.

Finally, in the concluding Sect.~V we summarize our results and point out several directions for future work.  In particular, we highlight two sets of open questions: the mathematical properties and physical consequences of traid and fraid group anyons and the conceptual and mathematical shifts required to formulate the quantum mechanics of indistinguishable particles on orbifolds.

\section{Orbifold approach to topological exchange statistics}

This section presumes that the reader has some familiarity with the topological approach to building a quantum theory on a configuration space $\cX$.  The key idea is that when a configuration space is not simply-connected, then single-valued wave functions defined on the configuration space may not exhaust the set of allowed states. The fundamental group $\pi_1(\cX)$ of the configuration space describes the connectivity of the $\cX$ by equivalence classes of loops. The irreducible representations of $\pi_1(\cX)$ provide a classification of possible wave functions, including single-valued and multi-valued as well as single-component (abelian) and multi-component (non-abelian).  Alternatively, one may work with only-single valued functions by lifting the quantum system to the unique, simply-connected universal cover of the configuration space $\widetilde{\cX}$.  For completeness, we provide a brief overview of these standard results for manifolds $\cX$ in Appendix A. We also provide a brief review of the main approaches to particle statistics, including exchange statistics and exclusion statistics, in Appendix B.

As first demonstrated by Leinaas and Myrheim~\cite{leinaas_theory_1977}, topological exchange statistics beyond FD or BE are possible when the base manifold upon which the particles move is not simply-connected or (in one and two dimensions) when particle interactions create topological defects.  They found the possibility for novel statistics by defining the configuration space of $N$ indistinguishable particles on a base manifold $\cM$ as the quotient of the configuration space manifold of $N$ distinguishable particles $\cX = \cM^N$ by the symmetric group $S_N$ of particle permutations~\cite{laidlaw_feynman_1971, leinaas_theory_1977}:
\begin{equation}\label{eq:Q}
  \cQ = \cX/S_N.
\end{equation}
Sometimes called the \emph{intrinsic} approach to indistinguishability~\cite{bourdeau_when_1992}, taking the quotient removes physically-meaningless particle labels from the mathematical description at the start. Although the taking the quotient precludes making passive permutations of particle identity, active particle exchanges are realized by loops in $\cQ$.

Taking the quotient also introduces a natural orbifold structure to $\cQ$.  Orbifolds have singular points where non-trivial local symmetries have topological consequences. Traditionally, these singular points have either been removed from configuration space or trivialized. Instead, we include these points and describe the particle exchanges of indistinguishable particles using the orbifold fundamental group $\pi_1^*(\cQ)$.

The generalization to $\pi^*_1$ captures the topological impact of singular points with co-dimension $\tilde{d}=1$ or $\tilde{d}=2$ on parallel transport.  Loci of singular points with $\tilde{d}=1$ are \emph{internal mirrors} and occur at two-body coincidences in one-dimensional systems.  Loci of singular points with $\tilde{d}=2$ are either isolated singularities, called \emph{cone points}, which occur in the case of particles in two dimensions, or \emph{corners} formed by intersections of $\tilde{d}=1$ intersections of internal mirrors and occur in the case of particles in one dimension. 

Using the orbifold fundamental group, we find the relation 
\begin{equation}\label{eq:gensym}
  \pi_1^*(\cQ) = S_N(\cM).
\end{equation}
holds for any path-connected base manifold $\cM$. Here $S_N(\cM)$ is the generalized symmetric group \cite{imbo_identical_1990}:
\begin{equation}\label{eq:gensym2}
  S_N(\cM) \equiv \pi_1(\cM)^N \rtimes S_N \equiv \pi_1(\cM) \wr S_N,
\end{equation}
where $\wr$ denotes the wreath product, a semidirect product in which $S_N$ acts on the normal subgroup $\pi_1(\cM)^N$ by permutation of factors \cite{james_representation_1984, harshman_one-dimensional_2016}. For a simply-connected base manifold, $S_N(\cM)$ reduces to the symmetric group $S_N$ as expected.

The result (\ref{eq:gensym}) was previously derived for $d = \dim\cM \geq 3$ in Ref.~\cite{imbo_identical_1990} where exchange statistics given by $S_N(\cM)$ are called generalized parastatistics. 
Extending the result (\ref{eq:gensym}) to particles on base manifolds $\cM$ with dimensions $d=1$ and $d=2$ requires the orbifold approach that we develop over the rest of this section.  The classification of topological exchange statistics in one dimension presented in Sect.~III can be understood without these details on orbifolds and orbifold fundamental groups.

\subsection{Distinguishable particle}

To understand the orbifold structure of the configuration space for indistinguishable particles, first consider \emph{distinguishable} particles moving on a path-connected manifold $\cM$ with $\dim{\cM} = d$.  The configuration space is
\begin{equation}\label{eq:X}
  \cX = \cM^N \equiv \overbrace{\cM \times \cdots \times\cM}^{N\ \mathrm{times}}.
\end{equation}
The fundamental group of $\cX$
\begin{equation}\label{eq:piX}
  \pi_1(\cX) = \pi_1(\cM)^N
\end{equation}
is the $N$-fold direct product of the fundamental group of $\cM$.  The universal cover $\widetilde{\cX}$ of $\cX$ factorizes similarly $\tilde{\cX} = (\widetilde{\cM} )^N$.

Removing particle coincidences may disrupt this product structure (\ref{eq:piX}) for base manifolds $\cM$ with dimension $d=1$ or $d=2$. To see this, note that every point in $\cX$ can be classified by its pattern of coinciding coordinates.  For \emph{generic} points $x\equiv \{x_1, \ldots, x_N\} \in \cX$, all $N$ coordinates are different. In contrast, the set of all points that have at least one pair of coordinates the same is called the coincidence locus $\Delta_2 \subset \cX$, or sometimes the `fat diagonal' of $\cX$~\cite{juhasz2018naturality}.  Similarly, one can define $\Delta_3 \subset \Delta_2$ as the locus where at least three particle coordinates are the same, $\Delta_{2,2} \subset \Delta_2$ as the locus where there are at least two pairs of coinciding particles, etc.

More generally, the space $\cX$ is stratified by integer partitions of $N$.  Each integer partition $[n_1\ldots n_k] \in P_N$ is a collection of positive integers which sum to $N$, typically written in non-increasing order.  For example, for $N=5$, the configuration space $\cX=\cM^5$ is stratified into seven partitions, including the partition $[221]$ corresponding to points with two $2$-particle coincidences and the partition $[41]$ for points with one $4$-particle coincidence.

For each partition $[\nu] = [n_1 \ldots n_k]$, we define the stratum $\cX_{[\nu]} \subset \cX$ of points with that partition type. The stratum $\cX_{[\nu]}$  has $h_{[\nu]} =  N!/(n_1!\cdots n_k!)$ path-connected components depending on which coordinates are equal. The closure of each stratum  $\overline{\cX_{[\nu]}}$ is defined by $\sum_{j=1}^k (n_j-1)$ equalities of $d$-dimensional variables giving us that $\cX_{[\nu]}$ is co-dimension $\tilde{d} = \sum_{j=1}^k (n_j-1) d$ within $\cX$.

The fat diagonal $\Delta_2 = \overline{\cX_{[21\ldots 1]}}$ is the union of all $\cX_{[\nu]}$ except the generic points in $\cX_{[1\ldots 1]}$ and its top-dimensional stratum has co-dimension $\tilde{d}=d$.  Therefore, for base manifolds $\cM$ with $d=1$ or $d=2$, removing $\Delta_2$ from $\cX$ disrupts the connectivity and the fundamental group is no longer given by (\ref{eq:piX}).  For $d=1$ base manifolds $\cM$, the coincidence loci $\Delta_3$ (the closure of $\cX_{[31\ldots 1]}$) and $\Delta_{2,2}$ (the closure of $\cX_{[2 2 1 \ldots 1]}$) have co-dimensions $\tilde{d}=2d=2$, as we discuss in detail Sect.~III.

As a final note, when the particles are distinguishable but identical, then particle permutations are a symmetry of the Hamiltonian.  There is a representation $O:S_N \to \Diff(\cX)$ denoted $s\mapsto O_s \in O(S_N)$ that acts as a \emph{passive} coordinate transformation, exchanging factors in the product (\ref{eq:X}).  Generic points $x \in \cX_{[1\ldots 1]}$ have orbits under permutation $\left\{O_s(x) \left| s \in S_N \right. \right\}$ with $N!$ members.  For all other partitions $[\nu]$, each point $x\in \cX_{[\nu]}$ has a non-trivial stabilizer subgroup $H_{x} \subset S_N$, i.e., the subgroup that exchanges coinciding coordinates.  Such points have orbits with only $N!/\left|H_{x}\right|$ members.  The particular embedding $S_{[\nu]} = S_{n_1}\times \cdots \times S_{n_k}$ into $S_N$ depends on which of the $h_{[\nu]}$ components of $\cX_{[\nu]}$ contains $x$, and two points of different components will will have conjugate stabilizers.

\subsection{Intrinsic approach}\label{subsec:intrinstic}

In the intrinsic approach, physics happens on the indistinguishable particle space $\cQ$ rather than $\cX$. The quotient $\cQ = \cX/S_N$ defines a `forgetful map' $p:\cX \to \cQ$ that erases the particles' identities and sends each point of $\cX$ to its $S_N$ orbit. For generic points in $\cX_{[1\ldots 1]}$, the map $p:\cX \to \mathcal{Q}$ is $N!:1$, but points in the coincidence locus $\Delta_2$ are \emph{singular} under this map. 

Reversing this, each possible choice of coordinate labels is equivalent to a `lift' of the point $q_0 \in \cQ$ to one of its representative points $x_0 \in \cX$ in the orbit.  Since particle labels are meaningless for indistinguishable particles, any assignment of labels to coordinates of $q_0$ is a choice of gauge. The diffeomorphisms $O_s$ of $\cX$ permute the representatives of $q_0$ and therefore the particle labels, so they form a discrete gauge transformation group. From another perspective, these label-permuting symmetries $O_s$ form a generalization of the deck transformations of a covering space in the orbifold category~\cite{Boileau_2003_Three}.

In contrast to the passive transformations $O_s$ of $\cX$, closed loops in $\cQ$ realize active particle exchanges along continuous paths.  However, paths that pass through the singular points $\Delta_2/S_N \subset \mathcal{Q}$ give an ambiguity.  Did the particles exchange at the coincidence point or not? Especially in one dimension, the ambiguity between reflection and transmission is essential to understanding exchange statistics.  However, the fundamental group $\pi_1(\mathcal{Q})$ does not see the path ambiguity embodied in these singular points~\cite{bourdeau_when_1992}.

There are two standard solutions to the presence of these singular points that avoid using orbifolds.  First, one can consider the space
\begin{equation}\label{eq:Q2}
 \cQ_2 = \cX_2/S_N = \cQ - \Delta_2/S_N,
\end{equation}
where $\cX_2 = \cX -\Delta_2$.  The configuration space $\mathcal{Q}_2$ has no multi-body coincidences and without these points, the restriction of $p$ that maps $\cX_2 \to \cQ_2$ forms a covering space in the usual sense.  Therefore $\cQ_2$ is a manifold and exchange paths are described completely by $\pi_1(\cQ_2)$.  Famously, for $\cM = \bR^2$, $\pi_1(\cQ_2)$ is the braid group $B_N$ and $\pi_1(\cX_2)$ is the pure braid group $PB_N$. On general $d=2$ surfaces $\cM$, the fundamental group $\pi_1(\cQ_2)$ gives the generalized braid groups $B_N(\cM)$ and pure braid groups $PB_N(\cM)$~\cite{birman_braid_1969, thouless_remarks_1985,  imbo_identical_1990, einarsson_fractional_1990, hatsugai_braid_1991}. For $\cM = \bR^3$, all particle exchanges which do not permute the particles are homotopically trivial and so $\pi_1(\cQ_2)= S_N$, resulting in normal exchange statistics.

Alternatively, one can consider the \emph{underlying space} $|\mathcal{Q}|$ of the orbifold $\cQ$, which includes set of singular points $|\Delta_2/S_N|$ but ignores their orbifold structure~\cite{Boileau_2003_Three}.  The space $|\cQ|$ is a manifold with certain degeneracies along $|\Delta_2/S_N|$ such as a boundary or corners.  In this case, one uses the usual fundamental group $\pi_1(|\cQ|)$ to describe particle exchanges.  However, trivializing the topology like this removes the possibility for non-trivial exchange statistics.  Returning to the case of $\mathcal{M} = \mathbb{R}^2$, we find $\pi_1(|\mathcal{Q}|) = 1$. This implies the unsatisfactory result that only trivial topological exchange statistics (i.e., only bosons, not even fermions) would be possible for particles moving on the plane unless interactions exclude $\Delta_2/S_N$~\cite{leinaas_theory_1977, wu_general_1984}.

\subsection{Configuration space orbifold}

Instead of either of these approaches, we include the singular points $\Delta_2/S_N$ in the configuration space and consider $\mathcal{Q}$ as an orbifold \cite{thurston_geometry_2002, Boileau_2003_Three}.  Orbifolds occur naturally in the context of classifying spaces of objects with symmetries \cite{adem_orbifolds_2007}.  The space $\cQ$ meets the definition of a `good' orbifold because it is the quotient of a manifold by a discrete group acting upon it by diffeomorphisms~\cite{Boileau_2003_Three}.  Further, $\cQ$ is considered `very good' as it actually is the global quotient of a manifold by a {\em finite} group. 

Points in an orbifold are classified as manifold points if their local symmetry group is trivial and orbifold points if their local symmetry group is non-trivial. Depending on their co-dimension, orbifold points form loci that appear as internal mirrors, cone points or corners, and other higher order singular points in the orbifold. Manifolds form a subclass of orbifolds where all points have trivial local symmetry groups. 

For the orbifold $\cQ$, the orbifold singular points are precisely the singular points $\Delta_2/S_N$ of $p: \cX \to \cQ$.  In the planar example, the orbifold locus of $\mathbb{R}^2/S_2$ has co-dimension $\tilde{d}=1$, forming an internal mirror edge at the line $\Delta_2/S_2$ of two-body coincidences. The local symmetry group of points on $\Delta_2/S_2 = p(\cX_{[2]})$ is isomorphic to $S_2$; all other points on the half plane $p(\cX_{[11]})$ are manifold points.

More generally, the local symmetry group of an orbifold point in $\cQ$ comes from its location in the stratification of $\cX$ by partitions, $\cX_{[\nu]}$. Each point of $\cX_{[\nu]}$ is invariant under a subgroup of $S_N$ isomorphic to $S_{[\nu]} = S_{n_1} \times \cdots \times S_{n_k}$. 
The stratification $\cX_{[\nu]}$ is invariant under the $S_N$ action so the decomposition descends to a similar stratification $p(\cX_{[\nu]})=\cQ_{[\nu]} \in \cQ$ where the stabilizers $S_{[\nu]}$ of points $x\in \cX_{[\nu]}$ become the local symmetry groups of the corresponding point $q=p(x)\in \cQ_{[\nu]}$. 

To understand how local symmetries act, first consider the case of a manifold point $q \in \cQ_{[1\ldots 1]}$. For a sufficiently small neighborhood $U$ of $q$ in $|\cQ|$ the preimage $p^{-1}(U)$ consists of $N!$ disjoint sets in $\cX$ all isomorphic to $U$ via $p$.  Each connected component $V$ of $p^{-1}(U)$ is a local model for $\cQ$ at $q$ and corresponds to an assignment of $N$ labels to the $N$ distinct points of $\cM$ defining $q$, or equivalently, an unambiguous ordering of those points in the product $\cX =\cM \times \cdots \cM$.  In other words, $p:\cX_{[1\ldots 1]} \to \cQ_{[1\ldots 1]}$ is a topological covering map and the local symmetry group of the point $q \in \cQ$ is trivial.

However, an orbifold point $q \in \cQ_{[\nu]}$, $\nu \neq [1\ldots 1]$, lifts to only $h_{[\nu]} = N!/|S_{[\nu]}|$ points in $\cX$.  Therefore, if $U$ is a small neighborhood of $q$ in $|\cQ|$, then $p^{-1}(U)$ will have $h_{[\nu]}$ connected components.  Although each connected component $V$ of $p^{-1}(U)$ serves as a local model for the orbifold $\cQ$ at $q$, they differ from the neighborhoods of manifold points in that they are not isomorphic to $U$.  Instead, each component $V$ of $p^{-1}(U)$ comes with an action of $S_{[\nu]}$ so that $U=V/S_{[\nu]}$.  The realization of the $S_{[\nu]}$ symmetries by particle-label permutations (i.e. the embedding $S_{[\nu]} \subset S_N$) depends non-trivially on the specific component $V$.  Each of the $V$ do not correspond to a canonical choice of particle labels, but instead corresponds to a choice of particle labels up to an $S_{[\nu]}$ ambiguity.

For example, take $N=3$ and consider points $q \in \cQ_{[2 1]}$. These points $q$ with the form $\{a,a,b\}$ have preimages $(a,a,b)$, $(a,b,a)$, and $(b,a,a)$ which have neighborhoods where the label-involution is realized by the permutations $(1,2)$, $(1,3)$, and $(2,3)$, respectively.  These are exactly the generators of the stabilizer subgroups of the respective points in $\cX_{[21]}$ and the resulting copies of $S_2\subset S_3$ are all conjugate.

\subsection{Orbifold fundamental group}\label{subsect:ofg}

The \emph{orbifold fundamental group} $\pi^*_1$ classifies based loops in an orbifold up to continuous deformation, generalizing the usual fundamental group \cite{thurston_geometry_2002, Boileau_2003_Three, harshman_anyons_2020}.  Intuitively, an orbifold path $\gamma: I \to \cQ$ consists of a map $|\gamma|: I \to |\cQ|$ of the underlying spaces together with compatible lifts to local models on the orbifold.  For example, if $\cQ = \mathbb{R}^2/S_2$, there are two types of paths which touch the singular boundary: one which lifts to the transmitted path in the local model and the other which lifts to the reflected path. These paths realize, and are distinguished by, the two elements of $\pi_1^*(\mathbb{R}^2/S_2)=S_2$.  As with $\pi_1$, choice of base point does not affect the isomorphism type of $\pi_1^*$ for path-connected spaces.

Note that for $d\geq 3$ the orbifold singular points have large enough co-dimension $\tilde{d} \geq 3$ that their presence or absence has no effect on the connectivity or the (orbifold) fundamental groups of $\cX$, $\cQ$, and $|\cQ|$.  Therefore, each of these (orbifold) fundamental groups are canonically isomorphic with $S_N(\mathcal{M})$.  However, for $d=1$ and $d=2$, the orbifold fundamental group `feels' the disruption to connectedness created by the local symmetry groups of orbifold loci.   Like the fundamental group, the orbifold fundamental group keeps track of how paths intersect with $\tilde{d}=1$ orbifold loci and wind around $\tilde{d}=2$ loci.

Using the orbifold fundamental group, the same relation $\pi_1^*(\cQ) = S_N(\cM)$ (\ref{eq:gensym}) holds for any dimension. When $\pi_1(\cM)$ is trivial, then $S_N(\cM)$ reduces to $S_N$ and the universal cover of $\cQ$ is the simply-connected space $\widetilde{\cQ} = \cX$ and topological exchange statistics reproduces the results of the symmetrization postulate in any dimension (see Appendix B).
For non-trivial $\pi_1(\cM)$, the universal covers $\widetilde{\cQ} = \widetilde{\cX} = \widetilde{\cM}^N$ coincide. The topology of the base manifold allows multiple exchange paths supporting the same particle permutation to be distinguished. Equivalently, there are paths that do not exchange particles that are not homotopically trivial.

These results can be summarized in a short exact sequence, a linear sequence of groups connected by homomorphisms such that the image of one homomorphism is the kernel of the next, and `short' in the sense that there are five group terms beginning and ending with the trivial group.  The following short exact sequence holds for any path-connected base manifold $\cM$ of any dimension and relates the fundamental group of $\cX$ to the orbifold fundamental group of $\cQ$:
\begin{subequations}\label{eq:ses}
\begin{equation}
  1 \to \pi_1(\cX) \to  \pi_1^*(\cQ) \to S_N \to 1, 
\end{equation}
which specifies to
\begin{equation}
1 \to \pi_1(\mathcal{M})^N \to  S_N(\mathcal{M}) \to S_N \to 1.
\end{equation}
\end{subequations}
The proof of (\ref{eq:gensym}) and (\ref{eq:ses}) in Ref.~\cite{imbo_identical_1990} holds when $d \geq 3$, but fails to generalize to the orbifold case.  However, we have an alternate proof that employs a distinct, but isomorphic definition of the orbifold fundamental group. In this definition, like the usual fundamental group, the orbifold fundamental group is defined to be the deck transformation group of the universal cover \cite{thurston_geometry_2002}.  Then the short exact sequence can be extracted from a series of covers $\widetilde{\cX} \to \cX \to \cQ$.  Here, the covering $\widetilde{\cX} \to \cX$ is the usual universal covering of the manifold $\cX$, which has deck transformation group isomorphic to $\pi_1(\cX)$ and the composition $\widetilde{\cX} \to \cQ$ is the universal cover of the orbifold $\cQ$.  Since $\cX \to \cQ$ comes from a group quotient, the action of the deck transformation group is transitive, and thus the cover is regular.  This means that $\pi_1(\cX)$ includes into $\pi_1^*(\cQ)$ as a normal subgroup with quotient equal to the deck transformation group $S_N$ of the cover $\cX \to \cQ$, giving the short exact sequence (\ref{eq:ses}).

However, we should warn the reader that we have done something sneaky here. When passing to the deck transformation interpretation of the the orbifold fundamental group, we have changed the object of study!  That is, for each cover, there are two groups at work here.  One is the group of passive transformations, acting as diffeomorphisms of the covering space, forming a discrete gauge group.  In the case of $\cX \to \cQ$, it is the group that permutes choices of particle labels and in the case of $\widetilde{\cQ}=\widetilde{\cX} \to \cQ$, it additionally intertwines how the constituent particles have wound around $\cM$.  The other group is the ``point-pushing'' group of orbifold-homotopy loops in the configuration space that realizes active transformations of indistinguishable particles.  These groups are isomorphic and that allows the proof of (\ref{eq:gensym}) and (\ref{eq:ses}), but the methods of action on $\widetilde{\cQ}$ are distinct and commute. In Sect.~\ref{subsect:interval} below, we contrast these actions for the simplest case of indistinguishable particles on a one-dimensional interval.

\section{Application to one dimension}

The orbifold fundamental group (\ref{eq:gensym}) of $\cQ$ describes how two factors determine the connectedness of configuration space: indistinguishability and the fundamental group of the base manifold.  In three dimensions and higher, those are the only two factors that contribute to topological exchange statistics.  However, hard-core or singular interactions exclude points from configuration space, and in $d=1$ and $d=2$, excluding points of coincidence alters $\pi_0$ (the set of path-connected components) and $\pi_1$ of configuration space because this locus has co-dimension $\tilde{d} \leq 2$.

A co-dimension $\tilde{d}=1$ defect, such as the set of two-body coincidences $\Delta_2$ in one dimension, locally splits configuration space into connected components.  Recall that $\Delta_2$ is formed as the union of sets defined by equations of the form $x_i=x_j$, so for a generic point, the local splitting is into two pieces in the same manner as a point on a line, a line in a plane, or a plane in a three-dimensional space.  When the intersections of all two-body coincidences $\Delta_2$ are removed to form $\cX_2 = \cX - \Delta_2$, particle exchanges through coincidences become impossible and the particles can be given a consistent order on each element of  $\pi_0(\cX_2)$. 

The removal of codimension $\tilde{d}=2$ defects, which sit in configuration space like a point in a plane or a line in a space, also disrupts the connectivity of configuration space.  Paths that wind around the defects lead to new, non-trivial elements of $\pi_1$ that serve as (often non-abelian) winding numbers.  Such defects occur as the result of the following few-body coincidences in $\cX$~\cite{harshman_anyons_2020, harshman_coincidence_2018}:
\begin{enumerate}
\item two-body coincidences $\Delta_2$ defined by $x_j = x_k$ and $y_j = y_k$ in two dimensions,
\item three-body coincidences $\Delta_3$ defined by $x_i = x_j = x_k$ in one dimension, and
\item a non-local, partial four body coincidences $\Delta_{2,2}$ formed by the intersection of two two-body coincidences $x_i = x_j$ and $x_k = x_l$.
\end{enumerate}
The first case famously leads to the braid group and generalization (see Sect.~\ref{subsec:intrinstic}).  We call (orbifold) fundamental groups that derive from the exclusion of these coincidence loci \emph{strand groups} because, like the braid group, they lead to configuration space described by generalizations of the symmetric group that can be realized by strand diagrams. However, we note that for the case of $\Delta_{2,2}$, this notion of strand group involves non-local constraints, as we discuss below.

Therefore, for $d=1$ base manifolds $\mathcal{M}$, we define the additional configuration spaces
\begin{eqnarray}\label{eq:tilded2}
  &\cX_3 = \mathcal{M}^N - \Delta_3, \cQ_3 = \cX_3/S_N \\
  &\cX_{2,2} = \mathcal{M}^N - \Delta_{2,2}, \cQ_{2,2} = \cX_{2,2}/S_N \nonumber\\
  &\cX_{\{3;2,2\}} = \mathcal{M}^N - \Delta_3 \cup \Delta_{2,2}, 
    \cQ_{\{3;2,2\}} = \cX_{\{3;2,2\}}/S_N \nonumber
\end{eqnarray}
where the spaces $\cX_3, \cQ_3$ excluding three-body coincidence hold interest for $N \geq 3$ and spaces excluding double two-body coincidences are relevant for $N \geq 4$.  Because these $\cQ$-spaces include single two-body coincidence orbifold locus $\Delta_2/S_N$, we consider them as orbifolds and use $\pi_1^*$.

There are only two homotopy types for manifolds in one dimension: the interval type with $\pi_1(\mathcal{M}) = 1$ (a type that includes the infinite interval $\mathbb{R}$) and the circle $S^1$ with $\pi_1(\mathcal{M}) = \mathbb{Z}$.  We classify the possible topological exchange statistics for both manifold types below for indistinguishable and distinguishable particles with and without the relevant coincident loci removed.

\subsection{Particles on interval-type manifolds}\label{subsect:interval}

When $\cM$ is of the simply connected interval type, the configuration space for distinguishable particles $\cX$ is also path-connected and simply-connected and there are no non-trivial exchange statistics.  For a finite interval, $\cX$ is an $N$-dimensional hypercube, but without loss of generality we extend to the infinite interval and consider $\cM = \bR$ and $\cX = \bR^N$.  If the particles are distinguishable but identical, then particle permutations are a symmetry of the Hamiltonian.  A permutation $s \in S_N$ is represented by an orthogonal transformation $O_s \in O(S_N) \subset O(N)$ on $\cX = \bR^N$; see Ref.~\cite{harshman_one-dimensional_2016a,harshman_one-dimensional_2016} for a pedagogical introduction.  The symmetrization postulate uses the representations $O(S_N)$ to decompose the Hilbert space of wave functions on $\cX$ into symmetric and antisymmetric subspaces (see Appendix B).

\begin{figure}
  \centering
  \includegraphics[width=\columnwidth]{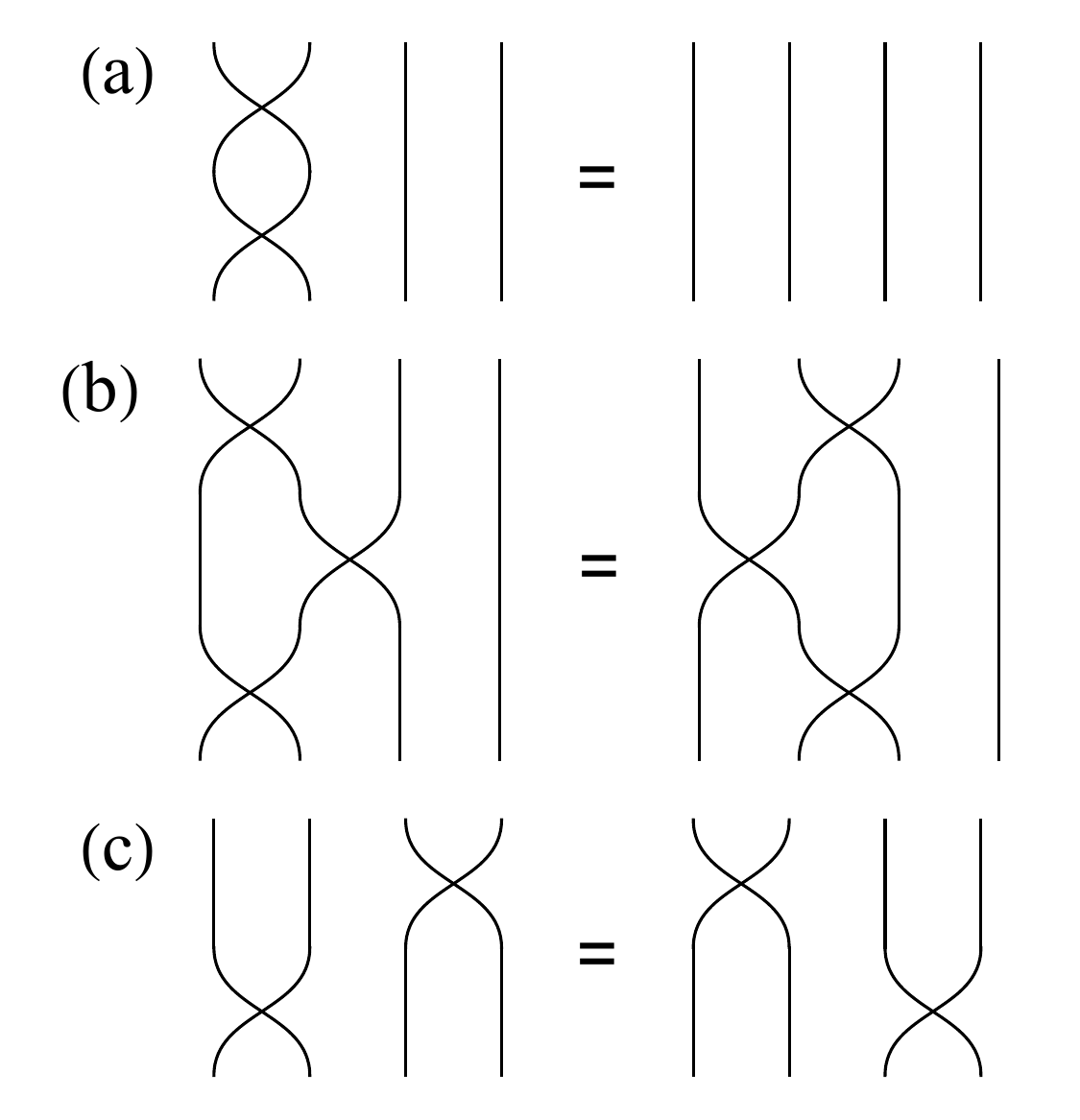}
  \caption{Depiction of the three types of generator relations (\ref{rel}) for $\pi_1^*(\cQ)$ when $N=4$ as strand diagrams, read from the bottom: (a) the self-inverse relation $\sigma_1^2 = 1$ that is broken for the braid group generators; (b) the Yang-Baxter relation $\sigma_1 \sigma_2 \sigma_1 = \sigma_2 \sigma_1 \sigma_2$ that is broken for the traid group generators; (c) the locality relation $\sigma_1 \sigma_3 = \sigma_3 \sigma_1 $ that is broken for the fraid group generators.}
  \label{fig:symrels}
\end{figure}

The intrinsic approach to indistinguishable particles, also gives the symmetric group $\pi_1^*(\cQ) = S_N$, as expected, but with a different interpretation in terms of exchange loops.  The orbifold fundamental group is generated by $N-1$ pairwise exchanges $\sigma_1$ through $\sigma_{N-1}$ satisfying the relations
\begin{subequations}\label{rel}
  \begin{eqnarray}
    \sigma_i^2 &=& 1 \label{rel:braid}\\
    \sigma_i \sigma_{i+1}\sigma_i &=& \sigma_{i+1} \sigma_i \sigma_{i+1} \label{rel:traid}\\
    \sigma_i \sigma_j &=& \sigma_j \sigma_i\ \mbox{for}\ j > i+1 \label{rel:fraid}.
  \end{eqnarray}
\end{subequations}
Recall that $\cQ$ is the space of indistinguishable particles, so the generators $\sigma_i \in \pi_1^*(\cQ)$ are not the discrete permutations of labeled particles. Instead, they are continuous paths that actively exchange particle \emph{orderings}, where each configuration of points has a natural particle order determined by the position of the particles in $\bR$.  The interpretation of $\sigma_1$ is that it realizes a loop in $\cQ$ that exchanges the first and second particle; $\sigma_2$ exchanges the second and third particles, etc.  These exchanges are realized as strand diagrams in Fig.~\ref{fig:symrels}.

\begin{figure*}
 \centering
 \includegraphics[width=\textwidth]{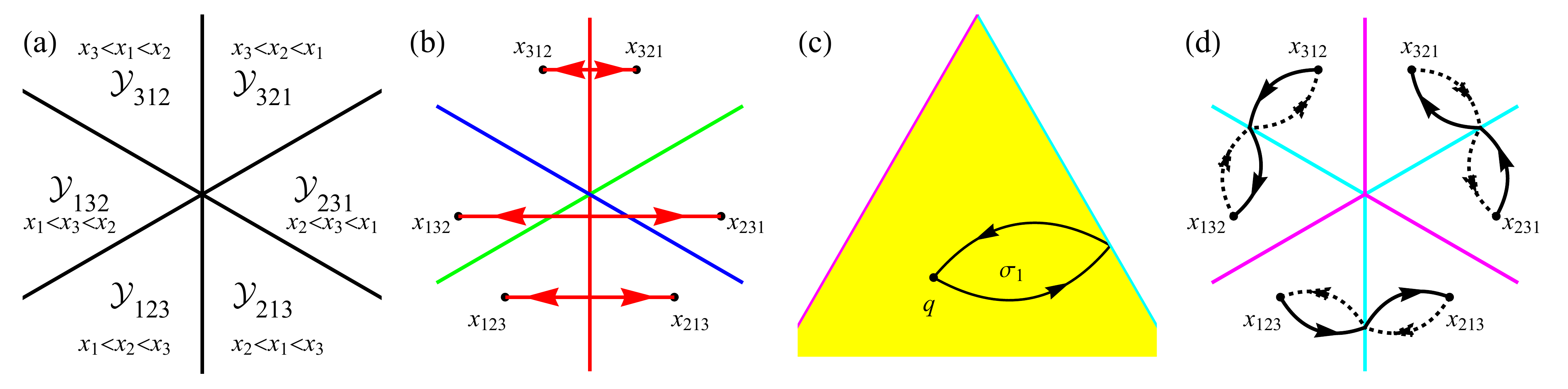}
 \caption{Graphical comparison of the actions of $O(S_N)\sim S_N$ and $\pi^*_1(\cQ) \sim S_N$ on $\cX_2$ for $N=3$. (a) The relative configuration space for three distinguishable particles on a line with coordinates $(x_1,x_2,x_3)$.  The horizontal coordinate is $z_1 = (x_1 - x_2)/\sqrt{2}$ and the vertical coordinate is $z_2= (x_1 + x_2 - x_3)/\sqrt{6}$. The black lines are the two-body coincidence locus $\Delta_2$ and they section $\cX$ into six alcoves $\cY_\omega \in \pi_0(\cX_2)$ where the particles have different position orders $x_{\omega_1} < x_{\omega_2} < x_{\omega_3}$.  (b) Reflections across each of the three colored lines are orthogonal transformations $O_s$ of $\cX$ that represent the passive permutation of particle labels.  For example, $O_{(12)}$ is a reflection across the vertical line (red) that permutes the labels of particle 1 and particle 2 and $O_{(23)}$ is a reflection across the diagonal line with positive slope (green) that permutes the labels of particles 2 and 3. The  double-sided arrows (red) connect points $x_\omega \in p^{-1}(q)$ and indicate how the $\cY_\omega \in \pi_0(\cX_2)$ are permuted by $O_{(12)}$.  (c) The relative components of the orbifold $\cQ$ and a based loop $\sigma_1 = \pi^*_1(\cQ)$ starting and ending at manifold point $q$. This path realizes an exchange $\sigma_1$ of the first two particles. (d) The based loop $\sigma_1$ in $\cQ$ is lifted to six paths in $\cX_2$ that start at $x_\omega$ and end at $x_{\omega'}$. For visual clarity, half of the path lifts are represented as dashed lines. These six lifts of element $\sigma_1 = \pi^*_1(\cQ)$ define a map $\tilde{\sigma}_1$ on $\cX_2$ that permutes the $\cY_\omega \in \pi_0(\cX_2)$.
 }
 \label{fig:deck}
\end{figure*}

We want to emphasize the difference between these two appearances of the symmetric group: (1) the active, continuous particle exchanges $\pi_1^*(\cQ) \sim S_N$ represented as closed loops on $\cQ$; and (2) the group of passive particle label permutations $O(S_N) \sim S_N$ represented as orthogonal transformations on $\cX$. We compare the action of these groups on the points of $\cX_2 = \cX - \Delta_2$ by restricting the action of $O(S_N)$ from $\cX$ to $\cX_2$ and lifting the action of $\pi_1^*(\cQ)$ from $\cQ$ to $\cX_2$. See Fig.~\ref{fig:deck} for a depiction of the these different actions for the case of $N=3$. 

\subsubsection{Singular two-body interactions}

For particles in one dimension with hard-core two-body interactions, the two-body coincidence locus $\Delta_2$ divides $\cX$ into $N!$ simply-connected alcoves $\cY_\omega \in \pi_0(\cX_2)$.
Each alcove $\cY_\omega$ is an open subset of $\cX$ such that the quotient map $p(\cY_\omega)$ consists entirely of manifold points $\cQ_2 \subset \cQ$. Each simply-connected component $\mathcal{Y}_\omega$ has the geometry of an open cone on an open simplex on the $N-1$ sphere~\cite{harshman_integrable_2017}. Conversely, each manifold point $q$ in $\cQ_2$ lifts to $N!$ points $p^{-1}(q)=\left\{x_\omega \mid x_\omega \in \cY_\omega, p(x_\omega)=q \right\}$.
Each alcove index $\omega = [\omega_1 \omega_2 \ldots \omega_N]$ is a permutation of the set $\{12\ldots N\}$ that indicates the positional ordering of labeled particles in $\bR$; see Fig.~\ref{fig:deck}.
When $\cM$ is of interval type, sorting in this manner gives a one-to-one correspondence between alcove orderings $\cY_\omega$ and permutations in $S_N$. However, as we explore in the next section, when $\cM=S^1$ we can only order particles cyclically and so this correspondence develops a similar cyclic ambiguity.

 For indistinguishable particles, $\mathcal{Q}_2$ is isomorphic to $\mathcal{Y}_\omega$ and the fundamental group $\pi_1(\cQ_2)=1$ is trivial. This provides a topological perspective on the so-called fermionization of hard-core bosons \cite{girardeau_relationship_1960, harshman_infinite_2017}. 
  Because no exchanges are possible, there are no topological exchange statistics that differentiate fermions and bosons from the intrinsic perspective.  As we discuss in Sect.~IV below, non-topological exchange statistics that interpolate between bosonic and fermionic solutions (sometimes called Leinaas-Myrheim anyons) have been defined by imposing Robin boundary conditions on $\Delta_2/S_N$ (equivalent to delta-interactions on $\Delta_2 \subset \cX$)  \cite{leinaas_theory_1977, posske_2017, balachandran_classical_1991}.

\subsubsection{Singular few-body interactions}

In contrast to $\cQ_2$, removing the $\tilde{d}=2$ few-body coincidences (\ref{eq:tilded2}) gives non-trivial tolopogical exchange statistics. The following strand groups are defined as:
\begin{subequations}\label{aidgroups}
  \begin{eqnarray}
    \pi^*_1( \mathcal{Q}_3) &=& T_N \\
    \pi^*_1( \mathcal{Q}_{2,2}) &=& F_N \\
    \pi^*_1( \mathcal{Q}_{\{3;2,2\}}) &=& W_N
  \end{eqnarray}
\end{subequations}
We call these discrete, infinite, non-abelian groups the traid group $T_N$ \cite{harshman_anyons_2020} (aka doodle group, planar braid group, twin group), the fraid group $F_N$, and the free Coxeter group $W_N$ (aka universal Coxeter group \cite{humphreys_reflection_1992}).

Like the braid group $B_N$, the strand groups $T_N$, $F_N$ and $W_N$ can be understood as resulting from eliminating generator relations of the symmetric group \cite{harshman_anyons_2020}.  Relaxing relation (\ref{rel:braid}), the generators $\sigma_i$ give the braid group $B_N$.  Relaxing (\ref{rel:traid}) gives the traid group $T_N$ and relaxing (\ref{rel:fraid}) gives the fraid group $F_N$.  Relaxing both (\ref{rel:traid}) and (\ref{rel:fraid}) means that only the self-inverse relations remain and the resulting group is the free Coxeter group $W_N$.  Each of these groups admit natural homomorphisms to $S_N$ obtained by reintroducing the lost relations.

These relations have a topological basis. When $d\geq 2$, each $\sigma_i$ can be represented by a path that avoids the coincidence locus $\Delta_2$ and the loci $\Delta_3$ and $\Delta_{2,2}$ are codimension $2d \geq 4$ and their removal does not affect the fundamental group as any null-homotopy of a loop can be arranged to avoid theses sets.  However for $d=2$, the relation $\sigma_i^2=e$ comes from a null-homotopy of a path which must pass through the codimension $2$ locus $\Delta_2$, so the presence of $\Delta_2$ determines the presence of the relation (\ref{rel:braid}).  In dimension $d=1$, paths representing $\sigma_i$ cannot be represented disjointly from $\Delta_2$.  However, the loci $\Delta_3$ and $\Delta_{2,2}$ are co-dimension $2d=2$ and the relations (\ref{rel:traid}) and (\ref{rel:fraid}) are induced by null-homotopies that must pass through these sets, respectively.  For example, any null-homotopy of the natural representative of $(\sigma_i \sigma_{i+1})^3$ must pass through have a strand diagram which has at least on one triple point. In other words, the null-homotopy passes through $\Delta_3$.

The pure version of each of these groups in (\ref{aidgroups}) is defined by looking at the corresponding distinguishable particle configuration space.  For example, we can define the pure traid as $PT_N = \pi_1(\cX_3)$. Following the argument of Sect.~\ref{subsect:ofg}, we construct the following short exact sequences analogous to (\ref{eq:ses}):
\begin{subequations}\label{paidgroups}
  \begin{eqnarray}
    &1 \to PT_N \to T_N \to S_N \to 1& \\
    &1 \to PF_N \to F_N \to S_N \to 1& \\
    &1 \to PW_N \to W_N \to S_N \to 1&.
  \end{eqnarray}
\end{subequations}
Some mathematical results for the pure traid group $PT_N$ can be found in~\cite{ bardakov_structural_2019, Naik20, Mostovoy20, MostovoyPresentation}.

The possible topological exchange statistics for these novel strand groups are classified by the irreducible representations of these groups.  The irreducible representations of $T_N$ and $PT_N$ have not received much attention from mathematicians, although the abelian representations of $T_N$ are classified and several non-abelian representations found in Ref.~\cite{harshman_anyons_2020}. However, there is still not a complete classification of the non-abelian irreducible representations of the braid group 75 years after the group was first described~\cite{artin_theory_1947}, so one should expect classifying the non-abelian representations of the traid and fraid groups will be a similarly complicated and long-standing mathematical project.

\subsection{Particles on rings}

For $\mathcal{M}=S^1$, the distinguishable particle configuration space $\cX = T^N$ is the $N$-torus $T^N = S^1 \times \cdots \times S^1$ with fundamental group $\pi_1(T^N) = \bZ^N$.  This abelian group of $N$-tuples of integers under addition describes equivalences classes of paths by how many times each particle winds around the ring.  It is generated by the $N$ translations $t_i$ of a single particle around the circle.  All irreducible representations of $\bZ^N$ are classified by $N$-tuples of phases $(\phi_1, \cdots, \phi_N)$ with $\phi_i\in [0, 2\pi)$.  Only for all $\phi_i=0$ are the wave functions on $\cX$ single-valued.  For identical but distinguishable particles, all phases take the same value $\phi_i = \phi$.

For indistinguishable particles, the orbifold fundamental group allows generalized parastatistics given by \begin{equation}\label{eq:Qring} \pi^*_1(\mathcal{Q}) = S_N(S^1) = \bZ \wr S_N = S_N \ltimes \bZ^N.
\end{equation}
This group is non-abelian even for $N=2$ and in addition to multi-valued, scalar representations~\cite{forte_quantum_1992}, there are multi-valued, multi-component wave functions on the orbifold $\mathcal{Q}$ that realize states with generalized parastatistics.  In principle, a complete classification of the irreducible representations of (\ref{eq:Qring}) can be found using the method of induced representations of the normal subgroup $\bZ^N$~\cite{altmann_induced_1977}.

\begin{figure}
 \centering
 \includegraphics[width=\columnwidth]{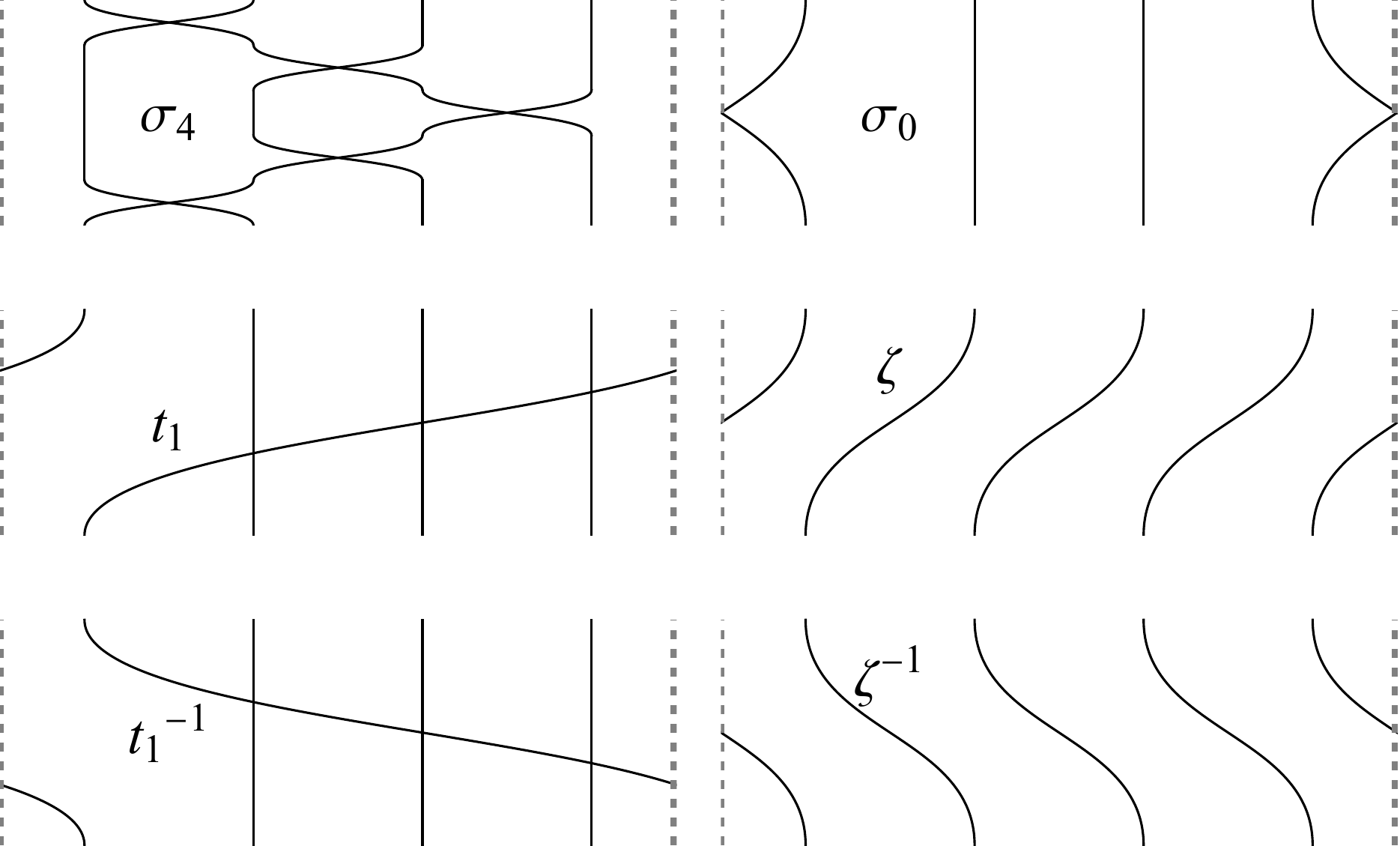}
 \caption{Depictions of six different elements of $S_4(S^1) = \bZ \wr S_4 = \bZ^4 \rtimes S_4$ mentioned in the text.  In these strand diagrams, the ring has been cut flattened into a line and the gray dashed lines on either side are the cut.  The elements depicted include: $\sigma_4$, the pairwise exchange of the first and fourth particle (relative to the cut) constructed from $S_N$ generators as in first line of (\ref{eq:sigmaN}); $t_1$, one of the $4$ generators of $\bZ^4$ subgroup of $S_4(S^1)$; its inverse $t_1^{-1}$; the `around-the-back' operator $\sigma_0$ defined in (\ref{eq:sigma0}) and simplified using $\sigma_i^2 = 1$; the $\zeta$ operators defined in (\ref{eq:zeta}) corresponding to a shift in the center-of-mass by $\pi/2$ around the ring; and $\zeta^{-1}$ its inverse.}
 \label{fig:affsym}
\end{figure}

\subsubsection{Twisted subgroup}

The group $\pi^*_1(\mathcal{Q}) = S_N(S^1)$ contains the affine symmetric group $\tilde{S}_N$~\cite{humphreys_reflection_1992} (also called the twisted symmetric group \cite{sutherland_beautiful_2004}) as a normal subgroup.  To show this, denote the $N-1$ generators of $S_N$ by $\sigma_i$ as in (\ref{rel}) and the $N$ generators of $\bZ^N$ by $t_i$.  The semidirect action of $S_N$ on $\bZ^N$ in the wreath product (\ref{eq:Qring}) implies the generator relations $t_i \sigma_i = \sigma_i t_{i+1}$ and $t_i \sigma_{j} = \sigma_{j} t_i$ for $j \neq i$ and $j \neq i +1$.  In terms of these generators, define the following three elements:
\begin{subequations}
  \begin{eqnarray}
    \sigma_N &=& \sigma_1 \sigma_2 \ldots \sigma_{N-2} \sigma_{N-1}\sigma_{N-2}\ldots \sigma_2 \sigma_1 \label{eq:sigmaN}\\
             &=& \sigma_{N-1} \sigma_{N-2} \ldots \sigma_2 \sigma_1\sigma_2\ldots \sigma_{N-2} \sigma_{N-1} \nonumber\\
    \sigma_0 &=& t_1 \sigma_N t_1^{-1} = t_1 t_N^{-1} \sigma_N \label{eq:sigma0}\\ 
    \zeta &=& t_1 \sigma_1 \sigma_2 \ldots \sigma_{N-1} = \sigma_1 \ldots \sigma_{N-1} t_N.\label{eq:zeta}
  \end{eqnarray}
\end{subequations}
These elements are depicted as strand diagrams in Fig.~\ref{fig:affsym}.  The element $\sigma_N$ is the pairwise exchange of the `first' and `last' particles (with respect to some starting angle on the ring) in which they pass through all the particles between them.  The element $\sigma_0$ is the pairwise exchange of the first and last particle `around the back', i.e., without crossing the other particles.  With the addition of $\sigma_0$, the set of elements $\{\sigma_0, \sigma_1,\ldots,\sigma_{N-1}\}$ satisfies the defining relations for generators of the affine symmetric group $\tilde{S}_N$~\cite{sutherland_beautiful_2004}.  This establishes that $\tilde{S}_N \subset S_N(S^1)$.

To show $\tilde{S}_N$ is a normal subgroup, consider the element $\zeta$.  It shifts all particles one place in the order and generates a subgroup $\bZ_\zeta \subset S_N(S^1)$ isomorphic to the integers.  It also satisfies the relation
\begin{equation}\label{eq:zetarel}
  \zeta \sigma_i = \sigma_{i+1}\zeta
\end{equation}
and in particular $\zeta \sigma_{N-1} = \sigma_0 \zeta$.  The element $\zeta \notin \tilde{S}_N$ acts as an outer automorphism on the generators of $\tilde{S}_N$ (and therefore on all of $\tilde{S}_N$) establishing  that $\tilde{S}_N$ is a normal subgroup of $S_N(S^1)$.  Therefore, the orbifold fundamental group can be equivalently expressed as
\begin{equation}\label{eq:Qringsep}
  S_N(S^1) = \bZ_\zeta \ltimes \tilde{S}_N
\end{equation}
where the semidirect product is specified by (\ref{eq:zetarel}).  Further note that $\zeta^N = t_1 t_2 \cdots t_N$ realizes a displacement of all particles one trip around the ring, i.e.~a full displacement of the center-of-mass.  One can show that $\zeta^N$ commutes with all elements of $\tilde{S}_N$.

\subsubsection{Singular two-body interactions}

Now consider $\cX_2$.  The removal of $\Delta_2$ divides the $N$-torus $T^N$ into only $(N-1)!$ sectors $\mathcal{Y}_\omega \in \pi_0(\cX_2)$ where $\omega$ labels a {\em cyclic} order of the particles.  There is one cyclic order for every coset $S_N/C_N$ of the symmetric group by the cyclic group $C_N \cong \bZ/N \equiv \bZ_N$.  For $N > 2$, hard-core two-body interactions lock the particles into a particular cyclic order $\omega$ and $\cX_2$ is not path-connected.  Spaces with multiple path components may, {\em a priori}, have non-isomorphic fundamental groups for each component.  However, within each ordering sector $\mathcal{Y}_\omega$, the fundamental group $\pi_1(\mathcal{Y}_\omega)$ is naturally isomorphic to the integers. We denote this group as $N \bZ_\zeta$ as it is generated by a full cycle around the ring by particles $\zeta^N$.  This ``full rotation'' interpretation also gives a natural isomorphism from $\pi_1(\mathcal{Y}_\omega)$ to the fundamental group of the base space, $\pi_1(S^1)$.

As in the interval case, $p: \mathcal{X}_2 \to \mathcal{Q}_2$ is not a connected cover, but unlike the interval case, $p$ is not simply an isomorphism from each $\mathcal{Y}_\omega$ to $\mathcal{Q}_2$.  Rather, the restriction of $p$ to $\mathcal{Y}_\omega$ is a connected $N:1$ cover.  Each point in $\mathcal{Q}_2$ corresponds to $N$ distinct points in $\mathcal{Y}_\omega$ that differ by a cyclic rotation $c \in C_N$ acting as a deck/gauge transformation of distinguishable particles.  The fundamental group $\pi_1(\mathcal{Q}_2)=\bZ_\zeta$ generated by $\zeta$ naturally contains $\pi_1(\mathcal{Y}_\omega) \sim \pi_1(S^1)$ as its $N \bZ_\zeta$ subgroup generated by $\zeta^N$.

These results establish that the corresponding short exact sequence of abelian groups for a connected component $\cY_\omega \subset \cX_2$ and $\cQ_2$ is
\begin{eqnarray}\label{eq:ses1}
  1 \to \pi_1(\cY_\omega) \to \pi_1(\cQ_2) \to C_N \to 1 \nonumber\\
  1 \to N\bZ_\zeta \to \bZ_\zeta \to \bZ/N \to 1.
\end{eqnarray}
Note that the symmetric group $S_N$ no longer appears in this sequence and therefore there are no FD statistics or parastatistics possible.  Instead the abelian group $\pi_1(\cQ_2)= \bZ_\zeta$ has one-dimensional representations characterized by a single phase $\phi$.

\begin{figure}
  \centering
  \includegraphics[width=\columnwidth]{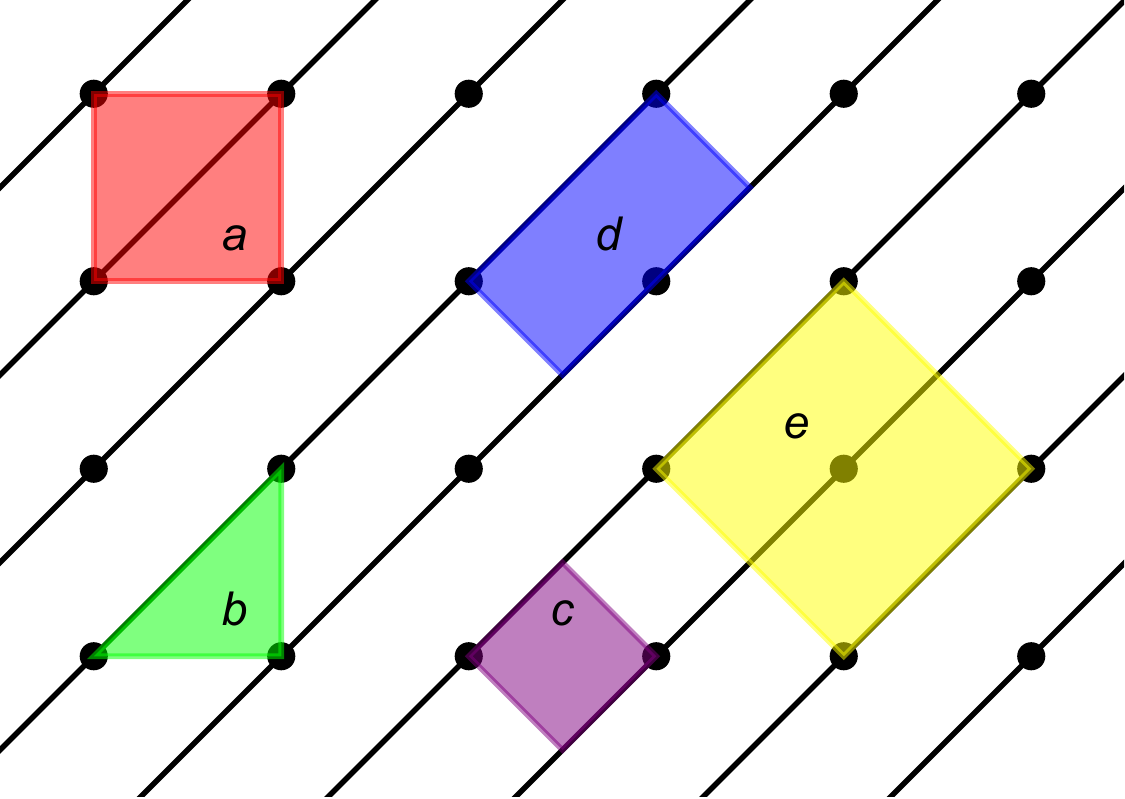}
  \caption{Depiction of the universal cover $\widetilde{\cQ}= \widetilde{\cX}$ and several useful domains for $N=2$ and $\cM=S^1$.  The black dots are all lifts of the same point in $\cQ$ and $\cX$, the black lines are the lifts of the coincidence locus $\Delta_2$ to $\widetilde{\cQ}$.  The red square (a) is one choice for the fundamental domain isomorphic to the torus $\cX = T^2$.  Each congruent square would then be labeled by a pair of integers in $\pi_1(\cX)=Z^2$ describing the path to fundamental domain.  The green triangle (b) and purple square (c) are equivalent alternate choices for the fundamental domain isomorphic to the M\"obius band $\cQ$ or $\cQ_2$.  The blue rectangle (d) is the double cover of $\cQ$ or $\cQ_2$.  It is the smallest torus $\overline{T}^2 = S^1_\mathrm{rel} \times S^1_\mathrm{com}$ for which there exists separable and single-valued center-of-mass and relative coordinates for indistinguishable particles. Although it has the same area in configuration space, it is not equivalent to the torus $\cX = T^2$.  The yellow square (e) is the double cover of $\cX$ that is the smallest torus for which exist single-valued center-of-mass and relative coordinates for distinguishable particles.}
  \label{fig:cover2}
\end{figure}

\begin{figure}
  \centering
  \includegraphics[width=\columnwidth]{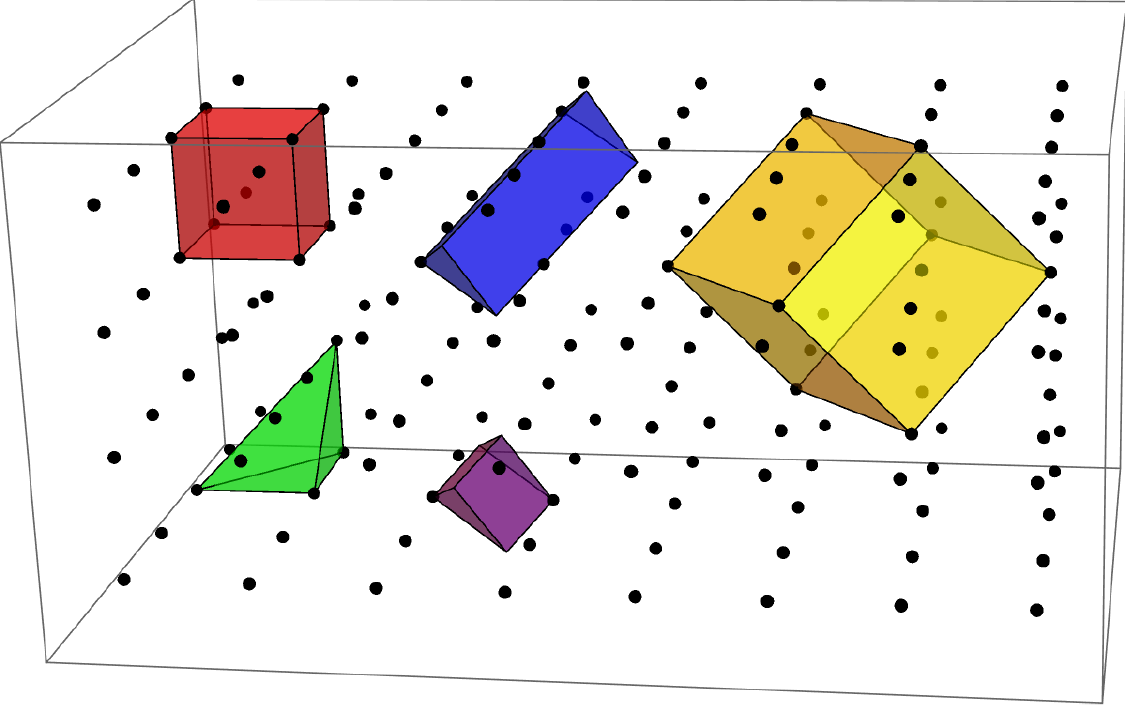}
  \includegraphics[width=\columnwidth]{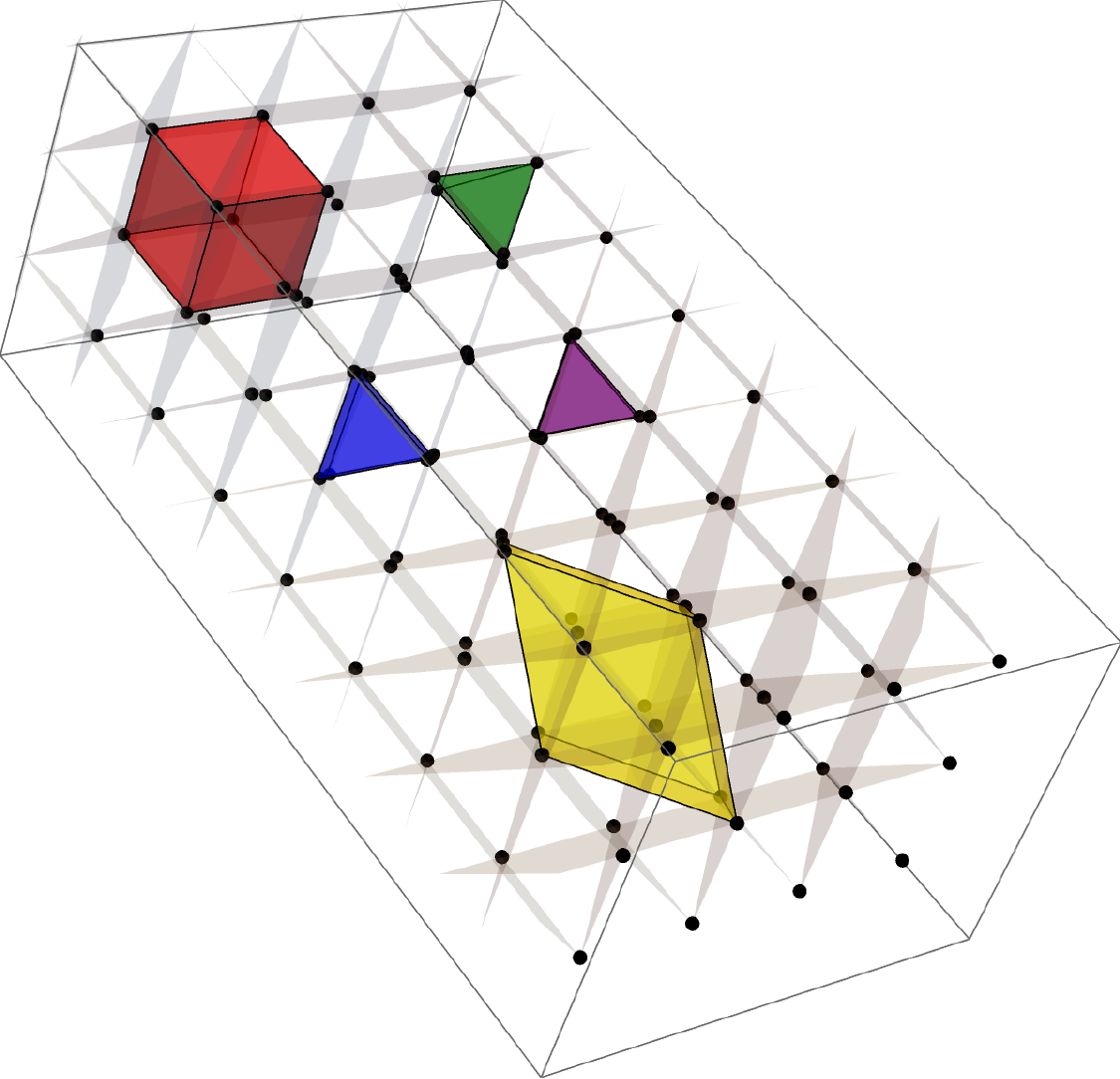}
  \caption{Two views of a depiction of the universal cover $\widetilde{\cQ}= \widetilde{\cX}$  and several useful domains for $N=3$ and $\cM=S^1$.  The black dots are all lifts of the same point in $\cQ$ and $\cX$, the black planes in the lower figure are the lifts of the coincidence locus $\Delta_2$ to $\widetilde{\cQ}$.  The five highlighted domains are equivalent to the five domains depicted in Fig.~\ref{fig:cover2}: the red cube is a fundamental domain congruent to $\cX = T^3$, the green 3-orthoscheme tetrahedron and purple triangular prism are equivalent alternate choices for the fundamental domain isomorphic to the Penrose triangle $\cQ$ or $\cQ_2$.  The blue rhombic prism is the minimal separable torus for indistinguishable particles $\overline{T}^3 = T^2_\mathrm{rel} \times S^1_\mathrm{com}$ and the yellow rhombic prism is the minimal separable torus for distinguishable particles.}
  \label{fig:cover3}
\end{figure}

The geometry of $\cQ_2$ provides insights into these results for $\pi_1(\cQ)$ and $\pi_1(\cQ_2)$.  The manifold $\cQ_2$ is a M\"obius band for the case $N=2$ \cite{leinaas_theory_1977}; see Fig.~\ref{fig:cover2}.  For $N=3$, the space $\cQ_2$ is a kind of Penrose triangle, a torus with an equilateral triangle cross section that makes a $2\pi/3$ twist every rotation; see Fig.~\ref{fig:cover3}. For higher dimensions, there is a generalized simplex hyperprism with a twist, e.g.~for $N=4$ the cross-section of the hyperprism is a rhombic dispheniod.  The ordering sector $\cY_\omega$ is an $N$-fold cover of $\mathcal{Q}_2$ that is a also a simplex hyperprism, but no longer twisted.  This universal cover of $\cQ_2$ is the product of $\bR$ and an $N$-simplex (non-regular for $N \geq 4$) and is useful for solving the Schr\"odinger equation in certain polytopes \cite{turner_quantum_1984, jain_exact_2008}.  For $\cQ$ the corresponding geometries are the same, but the orbifold singularities at the boundaries are included.

\subsubsection{Singular few-body interactions}

For co-dimension $\tilde{d}=2$ interactions, we define the following strand groups
\begin{subequations}\label{raidgroups}
  \begin{eqnarray}
    \pi^*_1( \mathcal{Q}_3) &=& T_N(S^1) \\
    \pi^*_1( \mathcal{Q}_{2,2}) &=& F_N(S^1) \\
    \pi^*_1( \mathcal{Q}_{\{3;2,2\}}) &=& W_N(S^1)
  \end{eqnarray}
\end{subequations}
Like the interval versions (\ref{aidgroups}), the equivalent groups (\ref{raidgroups}) result from breaking the analogous generator relations and each of these groups admit maps to $S_N(S^1)$ from relation reintroduction.

The factorizations (\ref{eq:Qring}) and (\ref{eq:Qringsep}) of $S_N(S^1)$ provides two alternate ways to view the groups in (\ref{raidgroups}).  For the case of $\pi^*_1( \mathcal{Q}_3)$, the two factorizations are
\begin{subequations}
  \begin{eqnarray}
    T_N(S^1) & = & T_N \ltimes \bZ^N \label{eq:traidwreath} \\
             &= & \bZ_\zeta \ltimes \tilde{T}_N \label{eq:twistedtraid}.
  \end{eqnarray}
\end{subequations}
In the first factorization, the normal subgroup $\bZ^N$ of particle translations around the ring is the same as (\ref{eq:Qring}), but $T_N$ replaces $S_N$.  The semidirect product in (\ref{eq:traidwreath}) obeys the same relations between the generators $\sigma_i$ and $t_j$ as (\ref{eq:Qring}).  Similarly, in the second factorization the `twisted' traid group $\tilde{T}_N$ is defined from $\tilde{S}_N$ by breaking the relation (\ref{rel:traid}) extended to the larger set of generators $\{\sigma_0, \ldots, \sigma_{N-1} \}$.  The semidirect product is inferred from the same generator relations as (\ref{eq:Qringsep}).

Equivalent constructions define the twisted fraid group $\tilde{F}_N$ and the twisted universal Coxeter group $\tilde{W}_N$.  In principle, these forms of the factorization provide a way to construct and classify irreducible representations of these novel strand groups on rings from the irreducible representations of strand groups on intervals.

Finally, pure versions of (\ref{raidgroups}) can be defined for the fundamental groups of the manifolds $\cX_3$, $\cX_{2,2}$ and $\mathcal{X}_{\{3;2,2\}}$ that satisfy short exact sequences like (\ref{paidgroups}).

\section{Topological analysis of models with one-dimensional anyons}

For some physicists, an abelian anyon model  must (by definition) have fractional exchange statistics, i.e., a pairwise exchange of particles transforms the wave function by a phase $\exp(i\theta)$ that interpolates between bosons $\theta=0$ and fermions $\theta = \pi$. Therefore, much previous work on anyons in one dimension either imposes fractional exchange statistics on wave functions (or field operators) at the start~\cite{ha_fractional_1995, zhu_topological_1996, girardeau_anyon-fermion_2006} or derives wave functions or field operators with fractional exchange phases from a Hamiltonian~\cite{kundu_exact_1999, keilmann_statistically_2011}.

However, in the previous section, we calculated the orbifold fundamental group (including the fundamental group in the case of $\cQ_2$) for the configuration space of indistinguishable particle for every possible topologically disruptive interaction on both intervals and rings.  For all scenarios, pairwise exchanges were either absent (as in $\cQ_2$) or square-trivial (for $\cQ$, $\cQ_3$, and the rest).  In no one-dimensional case is the (orbifold) fundamental group the braid group or any other group with abelian representations that furnish `traditional' fractional exchange statistics.

The necessary conclusion is that if fractional exchange statistics occur in models of particle systems in one dimension, they have a dynamical origin (i.e., they derive from interactions) and not a topological origin.  In other words, unlike in two dimensions, fractional exchange statistics do not derive from the configuration space of indistinguishable particles on one-dimensional manifolds even accounting for excluded few-body coincidences.  This in no way diminishes their physical or mathematical interest, but it may alter their interpretation and application.

To make the contrast clear, first consider the simplest case where `traditional' fractional exchange statistics given by a phase $\theta$ occurs: two particles in a plane with two-body coincidences excluded.  The configuration space $\cQ_2 = (\bR^4 - \Delta_2)/S_2$ can be factored into the product of a plane and cone with the point at the tip excluded~\cite{leinaas_theory_1977, bourdeau_when_1992}.  This space is not simply-connected and its fundamental group is $\pi_1(\cQ_2) = B_2 \cong \bZ$. The group $B_2$ therefore has abelian representations characterized by $\theta \in [0, 2\pi)$.  For $\theta \neq 0$, the wave function considered as a map $\psi: \cQ_2 \to \mathbb{C}$ is multi-valued.  Alternatively, one can define single-valued wave-functions the universal cover $\widetilde{\cQ_2}$, which the the product of a plane and a half-plane.  On the universal cover $\widetilde{\cQ_2}$ particle exchanges in $\pi_1(\cQ_2)$ are represented by translations.  The Hilbert space on $\widetilde{\cQ_2}$ decomposes into abelian representations of $\pi_1(\cQ_2)$ labeled by the quasi-momentum and $\theta$.

Besides working on $\cQ_2$ or $\widetilde{\cQ_2}$, there is a third option: to work with single-valued functions on $\cX_2$ and incorporate the $\theta$ exchange statistics as a gauge interaction potential for either bosons or fermions~\cite{khare_fractional_2005}.  This gauge potential is singular at the particle coincidence, but because that point is removed from configuration space $\cQ_2$ this singularity presents no difficulties.  For $N=2$, the gauge potential is the vector potential of a delta-function `flux tube' at the two-body coincidence~\cite{wilczek_magnetic_1982, wilczek_fractional_1990}.
This works for abelian braid group anyons because topological exchange statistics derive from flat connections on fiber bundles over configuration spaces. Therefore,  they can be absorbed into a gauge potential on a covering space, and a trivial gauge on the universal cover~\cite{ balachandran_classical_1991}.  As a result, free particles with fractional exchange statistics can be modeled with bosons (or fermions) where the statistics are `absorbed' into a statistical gauge interaction. 

This strategy of absorbing statistics into a gauge potential motivated Kundu~\cite{kundu_exact_1999} to define a one-dimensional anyon model (also called the anyon Lieb-Liniger model~\cite{posske_2017}). Starting from a bosonic model defined on $\cX$ with singular interactions described by $\delta$, $\delta'$ and double-$\delta$ functions on the $\Delta_2$ and $\Delta_3$ coincidence loci, Kundu performs a density-dependent gauge transformation~\footnote{Although discrete models are outside the scope of this analysis, a similar technique defines the anyon-Hubbard model~\cite{keilmann_statistically_2011}.  Density-dependent interactions for bosons are `transmuted' by the gauge transformation into site-dependent phase slips.  See also~\cite{bonkoff_bosonic_2021} for the relation of the anyon Hubbard model to the Kundu/Lieb-Liniger model.}.  The gauge-transformed model retains $\delta$-interactions on the $\Delta_2$ locus on which the creation and annihilation operators no longer satisfy bosonic relations.  The wave function that is constructed on $\cX$ jumps by $\exp(\pm i \theta)$ across $\Delta_2$.  Similarly, Girardeau defined a model with similar `phase slips' of $\exp(\pm i \theta)$ from a gauge transformation of a hard-core, fermion model which does not require singular interactions~\cite{girardeau_anyon-fermion_2006}.

The anyon Lieb-Liniger model has been shown to possess generalized \emph{exclusion} statistics~\cite{batchelor_one-dimensional_2006}. However, the phase slips $\exp(\pm i \theta)$ on $\cX$ in either the boson-based Lieb-Liniger model or fermion-based Girardeau model are the result of a particle-label dependent gauge transformation and do not have the same topological interpretation as fractional \emph{exchange} statistics of indistinguishable particles.  For example, in the simplest case of two particles, unlike the braid anyon case described above, there are not an infinite number of possible exchange phases multiplying the entire wave function depending on a winding number.  Instead, there is a single phase difference between the two different particle orderings of the same wave function defined on $\cX$.  The non-topological interpretation of Kundu/Lieb-Liniger anyons presented here agrees with several previous analyses that argue that fractional exchange statistics cannot be absorbed into a gauge potential for one-dimensional systems~\cite{aglietti_anyons_1996, valiente_bose-fermi-2020}. In contrast, because of their similar topological origin from excluded co-dimension $\tilde{d}=2$ coincidences, we hypothesize a `transmutation' from statistics to dynamics is possible for particles obeying traid or fraid exchange statistics, and is an avenue of future research.

Two other models sometimes identified as one-dimensional anyon models also have trivial topological exchange statistics: Leinaas-Myrhaim anyons and Calogero-Sutherland anyons.  In the Leinaas-Myrheim analysis of one-dimensional particles systems, the model is built on the exchange-trivial underlying space $|\cQ|$~\cite{leinaas_theory_1977}.  To make the Hamiltonian self-adjoint on the domain $|\cQ|$, boundary conditions must be imposed for the wave function on $\Delta_2$.  These self-adjoint extensions are characterized by an `anyon-like' parameter that effectively interpolates between Neumann boundary conditions with bosonic symmetry and Dirichlet boundary conditions with fermionic antisymmetry.  The Leinaas-Myrheim model can be lifted to the Lieb-Liniger bosonic model on $\cX$ with two-body delta-interactions, except the wave functions and all observables are restricted to the underlying space $|\cQ|$~\cite{posske_2017}. Like the Kundu model, here the interpolating statistics are again from the dynamics and not the topology, which is trivial. Further, we hypothesize that the Leinaas-Myrheim model on $|\cQ|$ can also be lifted to $\cX$ with fermionic antisymmetry or Kundu-like phase shifts using statistical mapping techniques~\cite{cheon_fermion-boson_1999, valiente_bose-fermi-2020, valiente_universal_2021, ohya_generalization_2021, ohya_discrete_2022}.

Similarly, in the Calogero-Sutherland model, interactions prevent two-body coincidences and a parameter interpolates between BE and FD exclusion statistics.  For indistinguishable particles, the inverse square interaction is sufficiently singular to exclude $\Delta_2$ and therefore the configuration space is $\cQ_2$.  Lifting the model to $\cX_2$, one is free to define arbitrary phases to different orderings of particles and define a Calogero-Sutherland \emph{anyon} model with fractional exclusion statistics and order-dependent phase slips \cite{ha_fractional_1995, polychronakos_generalized_1999, SREERANJANI20091176}.  However, since there is no exchange possible, such phases are a gauge symmetry that only has consequences for observables defined on the (non-universal) covering space $\cX_2$ and as such do not constitute a statistical gauge interaction.

\section{Conclusions}

To summarize, for indistinguishable particles on a one-dimensional interval we find the following possibilities for the group describing topological exchange statistics:
\begin{eqnarray}
  \pi_1^*(\cQ) &=& S_N \nonumber\\
  \pi_1(\cQ_2) &=& 1 \nonumber\\
  \pi_1^*(\cQ_3) &=& T_N \nonumber\\
  \pi_1^*(\cQ_{2,2}) &=& F_N \nonumber\\
  \pi_1^*(\cQ_{\{3;2,2\}}) &=& W_N.\label{interval}
\end{eqnarray}
The novel strand groups $T_N$, $F_N$ and $W_N$ result when three-body and certain four-body coincidences are removed from the configuration space orbifold $\cQ$.  Unlike the braid group, these co-dimension $\tilde{d}=2$ exclusions preserve the self-inverse property of pairwise exchanges (\ref{rel:braid}), but break the other defining relations of the symmetric group (\ref{rel:traid}) and (\ref{rel:fraid}). These groups provide the possibility of novel abelian and non-abelian anyons, but their irreducible  representations have not been classified or explored.

For indistinguishable particles on a circle, the underlying topology of the base space $S^1$ gets `mixed up' with the symmetric group in the same way the braid group on (for example) $S^2$ allows different topological exchange statistics than the more familiar braid group on $\mathbb{R}^2$. The equivalent groups to (\ref{interval}) expressed in two alternate forms are
\begin{eqnarray}
  \pi_1^*(\cQ) &=& \bZ^N \rtimes S_N = \bZ_\zeta \ltimes \tilde{S}_N \nonumber\\
  \pi_1(\cQ_2) &=& \bZ_\zeta \nonumber\\
  \pi_1^*(\cQ_3) &=& \bZ^N \rtimes T_N = \bZ_\zeta \ltimes \tilde{T}_N \nonumber\\
  \pi_1^*(\cQ_{2,2}) &=& \bZ^N \rtimes F_N = \bZ_\zeta \ltimes \tilde{F}_N \nonumber\\
  \pi_1^*(\cQ_{\{3;2,2\}}) &=& \bZ^N \rtimes W_N = \bZ_\zeta \ltimes \tilde{W}_N.
\end{eqnarray}
 The affine strand groups $\tilde{T}_N$, $\tilde{F}_N$ and $\tilde{W}_N$ arise from broken relations of $\tilde{S}_N$ in the same way as the non-affine versions. As far as we know, this is the first time these hyperbolic groups have been identified in a physical system and their group structures and irreducible representations are largely unexplored. Also, except for the case of $\cQ_2$, all of these groups provide the possibility for novel abelian and non-abelian anyons.

In none of these cases does the group giving topological exchange statistics furnish an abelian representation with fractional exchange statistics for an arbitrary $\theta$.  Our conclusion agrees with \cite{aglietti_anyons_1996, valiente_bose-fermi-2020} that if a one-dimensional model exhibits fractional exchange statistics, these are of a dynamical origin and cannot be absorbed into a consistent gauge potential.  In contrast, because they have a topological origin, the alternate strand groups like $T_N$ and $\tilde{T}_N$ described above should be `transmutable' into a gauge potential, and this looks to be a promising avenue for investigating there phenomenological signatures.  Because the traid and fraid group also derive from co-dimension two topological defects in configuration space, we hypothesize there could be similarities to braid group anyons where connections among conformal field theory, fusion rules, and quantum groups is a productive line of research; c.f.~\cite{ALVAREZGAUME1990347}. A field theoretical formulation of these alternate strand groups is certainly required for applications to many-body systems.

Are these alternate strand groups feasible to realize physically? Ultracold atoms can be confined to effectively one-dimensional traps, and in principle hard-core three body interactions can be engineered in cold atoms systems \cite{buchler_three-body_2007, daley_effective_2014, mahmud_dynamically_2014, valiente_three-body_2019}.  Density-induced interactions in Floquet-driven lattice models also show promise~\cite{greschner_2015, straeter_2016}. Hard-core three body interactions would effective exclude $\Delta_3$ from $\mathcal{Q}$, and dynamical models with three body interactions have shown signs of novel statistics and other thermodynamic properties \cite{paredes_pfaffian-like_2007, keilmann_statistically_2011, sowinski_criticality_2015, arcila-forero_three-body-interaction_2018}.  In Ref.~\cite{harshman_anyons_2020}, we have described and depicted the lowest energy wave functions obeying abelian traid exchange statistics for three particles in a harmonic trap, but exploring these solutions for more particles and classifying how they transform under discrete symmetries is an ongoing project.

In contrast, the paired two-body interactions necessary to exclude $\Delta_{2,2}$ are non-local.  A single two-body coincidence would need to prevent other two-body coincidences anywhere in the system.  As a fundamental interaction, non-local interactions are typically excluded, but as an effective theory for long-range interactions in a many-body system, it may have interest. More generally, because these novel strands groups arise `naturally' as degenerations of the ubiquitous symmetric group, one can image that they could emerge in a variety of non-particle model contexts. 

As a final note, building an `intrinsic' quantum theory for indistinguishable particles directly on the quotient space, without reference to a covering space of distinguishable particles, is an incomplete project that requires mathematical and conceptual definitions.  Operators that are self-adjoint on $\cX$, such as the Hamiltonian and the single-particle position and momentum, no longer have that property when restricted to $\cQ$~\cite{bourdeau_when_1992, balachandran_classical_1991, Gaveau_2012}.  For relative momentum and other operators that are not symmetric under particle exchange, self-adjointness cannot be restored and their interpretation does not seem to extend unambiguously to indistinguishable particles.

\acknowledgements{We would like to thank Franscesca Ark, Andr\'e Eckardt, Andreas Bock Michelsen, Zachary Tomares, Philip Johnson, Maxim Olshanii, Thore Posske, and Thomas Schmidt for useful discussions.}

\appendix

\section{Topological quantization}

The topological approach to quantization developed from several independent directions~\cite{schulman_path_1968, laidlaw_feynman_1971, dowker_quantum_1972, leinaas_theory_1977, mead_determination_1979, goldin_representations_1981, wilczek_quantum_1982,Isham:1983zr, berry_quantal_1984}. Possible quantizations of a configuration space $\cX$ are classified using irreducible unitary representations of the fundamental group $\pi_1(\cX)$.  The fundamental group $\pi_1(\cX)$ describes equivalence classes of based loops on on a connected manifold $\cX$.  Whenever $\cX$ is connected, the isomorphism type of $\pi_1(\cX)$ is independent of the base point.

Abelian representations of $\pi_1(\cX)$ characterize flat $U(1)$ connections on flat line bundles over $\cX$ and record the holonomy of the scalar wave function around closed loops. For non-trivial representations, the corresponding wave functions transform by a phase under holonomy and are therefore considered multi-valued wave functions on $\cX$, also called $\theta$-structures~\cite{Isham:1983zr, imbo_inequivalent_1988}.
Similarly, non-abelian representations, considered up to conjugation, characterize flat $U(\mu)$ connections on higher-rank fiber bundles leading to multi-valued, multi-component wave functions~\cite{imbo_identical_1990, balachandran_classical_1991, lee_introduction_2018}.

If working with multi-valued wave functions on $\cX$ is distasteful or inconvenient, one can instead work with single-valued wave functions on the simply-connected universal cover of $\cX$, denoted $\widetilde{\cX}$ with covering map $p_\cX: \widetilde{\cX}\to \mathcal{X}$.
This covering map also defines the original configuration space as a quotient space
\begin{equation}
  \cX = \widetilde{\cX}/D
\end{equation}
where $D$ is a group of deck transformations (diffeomorphisms on $\widetilde{\cX}$ compatible with the covering map $p_\cX$) isomorphic to $\pi_1(\cX)$. Some basic examples include the universal cover of a circle $S^1$ by the real line $\bR$ and the universal cover of the $n$-torus by $\bR^n$.  Another famous example is the quantum rotor, whose configuration space $\mathrm{SO}(3)$ is not simply-connected.  The fundamental group $\pi_1(\mathrm{SO}(3)) = Z_2$ has two irreducible representations.  The non-trivial irreducible representations corresponds to wave functions with half-integer spin that are double-valued on $\mathrm{SO}(3)$, but single-valued on its universal cover $\mathrm{SU}(2)$~\cite{schulman_path_1968}.  In these cases, working with single-valued functions on the universal cover can be simpler than multi-valued functions on the base manifold, but that is certainly not always the case.  For example, the universal cover of the configuration space whose fundamental group is the braid group $B_N$ is quite complicated for $N \geq 3$.

\section{Particle statistics in quantum mechanics}

Two main approaches to characterizing particle statistics are \emph{exchange statistics} and \emph{exclusion statistics}.  Exchange statistics considers how wave functions or operators transform when indistinguishable particles are exchanged.  Further, there are two main approaches to exchange statistics in quantum mechanical systems: the \emph{symmetrization postulate} and \emph{topological exchange statistics}, and we briefly compare them in this appendix. We focus on systems of particles on manifolds and do not consider the interesting case of particles on graphs~\cite{harrison_quantum_2011, harrison_n-particle_2014, maciazek_non-abelian_2019} which may have novel applications to tight-binding lattice models. We also restrict ourselves to quantum mechanics in configuration spaces and Hilbert spaces with a fixed number of particles and do not consider the important case of field operators and Fock spaces. For completeness, exclusion statistics is briefly described at the end of this appendix.

The first and oldest approach to exchange statistics of indistinguishable particles in quantum mechanics is the symmetrization postulate.  For the sake of simplicity, consider a single-component, single-particle Hilbert space $\cH = L^2(\cM)$ of square-integrable wave functions on the manifold $\cM$. 
For $N$ distinguishable but identical particles, the total Hilbert space $\cH^N$ can be constructed as the $N$-fold tensor product $\cH^N = \cH \otimes \cdots \otimes \cH$. This space is equivalent to the Hilbert space $\cH^N = L^2(\cX)$ of square-integrable functions on the total configuration space $\cX = \cM^N$. Note that the symmetrization postulate also applies to multi-component single-particle wave functions living in $\cH = L^2(\cM) \otimes \mathbb{C}^\mu$; see \cite{harshman_one-dimensional_2016} for examples with spin components.

Indistinguishability is imposed by symmetrizing a system of identical distinguishable particles. This is accomplished using representations of particle permutations $s\in S_N$ as unitary operators $U(s)$ that act on $\cH^N$. The action of operators $U(s)$ on  $\cH^N$ is induced from the realization of passive particle permutations $s\in S_N$ as diffeomorphisms $O_s$ acting on $\cX$. Note that when $\cM$ is an interval in $\bR$ then $\cX$ is a box in $\bR^d$ and the diffeomorphisms are a representation of $s\in S_N$ by orthogonal matrices $O_s \in \mathrm{O}(Nd)$.

The two abelian representations of $S_N$  provide the technology relevant to constructing the state space for bosons and fermions. Wave functions obeying BE exchange statistics are elements of $\cH^N_+$, the projection of $\cH^N$ onto the symmetric subspace, and wave functions in the antisymmetric subspace $\cH^N_-$ have FD exchange statistics.  After symmetrization, each choice of particle labels is equivalent to a choice of gauge within $\cH^N_\pm$~\cite{polychronakos_generalized_1999, bourdeau_when_1992}.

The symmetrization postulate can also be extended to the non-abelian irreducible representations of $S_N$ for $N \geq 3$.  Wave functions belonging to these representations obey so-called parastatistics~\cite{green_generalized_1953, messiah_symmetrization_1964}. These multi-component wave functions are useful for constructing states of indistinguishable fermions or bosons that have internal degrees of freedom~\cite{sudarshan_configuration_1988, polychronakos_generalized_1999, harshman_one-dimensional_2016}.

However, the symmetrization postulate approach to indistinguishability presumes single-valued wave functions on $\cX$ with $S_N$ exchange statistics.  As we discuss in Sect.~IV, non-$S_N$ exchange statistics can be mimicked by adding interactions to fermions or bosons.  Alternatively, multi-valued wave functions with non-$S_N$ statistics arise `kinematically' (i.e., without interactions) in the topological approach to exchange statistics when the configuration space is not simply-connected~\cite{leinaas_theory_1977, balachandran_classical_1991}.  Instead of treating exchanges as global transformations, exchanges are specific active continuous processes \cite{messiah_symmetrization_1964} which can be visualized as strand diagrams, i.e.\ $N$ paths on $\cM$ that connect particle configurations differing by at most a permutation of identical particles.  When the $N$ paths do not coincide on $\cM$, they can be lifted unambiguously to one path in $\cM^N$, up to a label permutation gauge.

Topological exchange statistics also begins from the configuration space $\cX$ for $N$ distinguishable particles on a connected manifold $\cM$ (\ref{eq:X}).  If the base manifold has dimension $\dim\cM = d$ and connectivity described by the fundamental group $\pi_1(\cM)$, then $\cX$ is a manifold with dimension $\dim\cX = Nd$ and fundamental group $\pi_1(\cX)= \pi_1(\cM)^N$.

On $\cX$, paths that start and end at the same place realize particle `exchanges' where every particle returns to its original location, but if $\cM$ is not simply connected, then even these paths can be topologically non-trivial and wave functions on $\cM$ need not be single-valued.  For each irreducible representation of $\pi_1(\cM)^N$ there is a class of multi-valued solutions.  If the particles on $\cM$ are distinguishable but \emph{identical}, then one can restrict to a diagonal representation of $\pi_1(\cM)^N$ that factors into the $N$-fold product of the same irreducible representation of $\pi_1(\cM)$.

The configuration space for indistinguishable particles, $\cQ = \cX/S_N$, is the topological quotient of $\cX$ by the symmetric group~\cite{laidlaw_feynman_1971, leinaas_theory_1977}.  However, as described in Sect.~II, the space $\cQ$ has singular orbifold points with a non-trivial topological structure. The standard solution removes these singular points (and therefore the path ambiguity) from configuration space entirely \cite{laidlaw_feynman_1971, leinaas_theory_1977, bourdeau_when_1992, wu_general_1984, nayak_non-abelian_2008}. The motivation for the orbifold approach is to resolve this ambiguity.

The orbifold fundamental group allows a derivation of symmetric group statistics (i.e., bosons, fermions and parastatistics) for simply-connected base spaces and generalized parastatistics for multiply-connected base spaces in a unified fashion and without needing to exclude path-ambiguous two-body collisions. For fermions and bosons on simply-connected manifolds with non-singular interactions, the orbifold approach to topological exchange statistics gives the same results as the symmetrization postulate in any dimension, whereas the standard approach to topological exchange statistics does not for $d \leq 2$ without introducing additional constraints or dynamics~\cite{leinaas_theory_1977, bourdeau_when_1992}.

Finally, in contrast to either form of exchange statistics, exclusion statistics~\cite{haldane_``fractional_1991, POLYCHRONAKOS1996202} is a more recent approach to understanding particle statistics. It employs a fractional statistics parameter $g$ to describe how the effective dimension of the single particle Hilbert space changes as the number of particles increases~\footnote{A related notion is Gentile statistics characterized by the maximum occupation number of a state~\cite{gentile_osservazioni_1940, polychronakos_generalized_1999, dai_gentile_2004}.}.  This parameter is $g=0$ for BE statistics because any number of bosons can occupy the same state, and $g=1$ for FD statistics because the Pauli exclusion principle requires an additional state for each fermion.  This parameter is consistent with the fractional exchange statistics parameter for abelian representations of the braid group~\cite{johnson_haldane_1994}.  Unlike the topological approach to exchange statistics, formulating indistinguishability in terms of exclusion statistics does not depend on the dimension of the underlying physical space $\cM$.  As a result, it has been especially useful for analyzing one-dimensional dynamical models \cite{ha_fractional_1995, batchelor_one-dimensional_2006, murthy_exclusion_2013} for which the standard (non-orbifold) topological approach gives trivial results.

\bibliographystyle{apsrev4-2}

\end{document}